\documentclass[12pt]{article}
\usepackage{amsthm,latexsym,multibox,amssymb,amsfonts,graphicx,cite}
\usepackage{xy}
\xyoption{all}
\input amssym.def
\input amssym.tex

\newcommand{\ft}{\footnotesize}
\def\ie{{\it i.e. }}
\def\a{\alpha}
\def\b{\beta}
\def\g{\gamma}
\def\d{\delta}

\def\h{\eta}
\def\l{\lambda}

\def\m{\mu}
\def\n{\nu}
\def\r{\rho}
\def\o{\omega}
\def\s{\sigma}

\def\t{\tau}

\def\ve{\varepsilon}

\def\pa{\partial}

\def\be{\begin{equation}}
\def\ee{\end{equation}}
\def\bqn{\begin{eqnarray}}
\def\eqn{\end{eqnarray}}
\def\nn{\nonumber}

\def\cl{{\cal L}}

\def\co{{\cal O}}

\newtheorem{theorem}{Theorem}[section]
\newtheorem{lemma}{Lemma}[section]

\usepackage{color}
\definecolor{rougef}{rgb}{0.56,0,0}
\definecolor{vertf}{rgb}{0,0.5,0}
\definecolor{bleuf}{rgb}{0,0,0.8}

\newlength{\blength}
\renewcommand{\proof}[1]{\vspace{-.05cm}
\begin{list}{\bf Proof:}
{\listparindent=\parindent\parsep=0pt \labelwidth=-0.5cm
\labelsep=\parindent \addtolength{\labelsep}{-\blength}
\addtolength{\labelsep}{1.5cm} \itemindent=-\blength
\addtolength{\itemindent}{\parindent} \leftmargin=1.0cm} \item
#1~$\qedsymbol$\end{list} \vspace{.0cm}}

\begin{document}

\begin{titlepage}
 \begin{flushright}
ULB-TH/04-21 \\
DAMTP-2004-67\\
DFPD/04/TH/14
 \end{flushright}
 \vskip 2cm

 \begin{centering}

 {\large {\bf No Self-Interaction for Two-Column Massless Fields }}

 \vspace{2cm}
Xavier Bekaert$^{a}$,
 Nicolas Boulanger$^{b}$ and
 Sandrine Cnockaert$^{c,}\footnote{``Aspirant du F.N.R.S., Belgium''}$ \\
 \vspace{1.5cm}
 {\small
$^a$ Dipartimento di Fisica, Universit\`a degli Studi di Padova, INFN, Sezione di Padova\\
Via F. Marzolo 8, 35131 Padova, Italy\\
\vspace{.2cm}
 $^b$ Department of Applied Mathematics and Theoretical Physics\\
 Wilberforce Road, Cambridge CB3 0WA, UK\\
 \vspace{.2cm}
 $^c$ Physique Th\'eorique et Math\'ematique and
International Solvay Institutes,\\
 Universit\'e Libre
 de Bruxelles,  C.P. 231, Bld du Triomphe, 1050 Bruxelles, Belgium }     \\

\vspace{.7cm}
{\tt \footnotesize xavier.bekaert@pd.infn.it, N.Boulanger@damtp.cam.ac.uk, \\
 Sandrine.Cnockaert@ulb.ac.be}

\vspace{1.5cm}

\end{centering}

\begin{abstract}
We investigate the problem of introducing consistent
self-couplings in free theories for mixed tensor gauge fields
whose symmetry properties are characterized by Young diagrams made
of two columns of arbitrary (but different) lengths. We prove
that, in flat space, these theories admit no local,
Poincar\'e-invariant, smooth, self-interacting deformation with at
most two derivatives in the Lagrangian. Relaxing the derivative
and Lorentz-invariance assumptions, there still is no
deformation that modifies the gauge algebra, and in most cases no
deformation that alters the gauge transformations.
 Our approach is based
on a BRST-cohomology deformation procedure.
 \end{abstract}

 \vfill
 \end{titlepage}


 \section{Introduction}

These last few years, mixed symmetry gauge fields (\ie that are
neither completely symmetric nor  antisymmetric) have
attracted some renewed attention
\cite{Brink:2000ag,Burdik:2001hj,Hull:2001iu,Bekaert:2002dt,deMedeiros:2002ge,
Boulanger:2003vs,Bekaert:2003zq,Alkalaev:2003qv,Alkalaev:2003hc,deMedeiros:2003px},
thereby reviving the efforts made in this direction during the
eighties, under the prompt of string field theory
\cite{Curtright:1980yk,Aulakh:cb,Siegel:1986zi,Labastida:1986gy}.
Mixed-symmetry fields appear in a wide variety of
higher-dimensional ($D>4$) contexts. Indeed, group theory imposes that
first-quantized particles propagating in flat background should
provide  representations of the Poincar\'e group. The cases $D=3\,,4$ are very particular in the sense that
each tensor irreducible representation (irrep.) of the little
groups $O(2)$ and $O(3)$ is equivalent to a completely symmetric
tensor irrep. (pictured by a one-row Young diagram with $S$
columns for a spin-$S$ particle). When $D>4\,$, more complicated
Young diagrams are allowed. For instance, all critical string
theory spectra contain massive fields in mixed symmetry representations
of the Lorentz group. In the tensionless limit
($\alpha^\prime\rightarrow \infty$) all these massive excitations
become massless. Another way to generate various mixed symmetry
fields is by dualizing totally symmetric fields in
higher dimensions \cite{Hull:2001iu,Boulanger:2003vs}. \vspace{.4cm}

An irrep. of the general linear group
$GL(D,\mathbb{R})$ is denoted by $[c_1,c_2,\ldots,c_L]\,$, where
$c_i$ indicates the number of boxes in the $i$-th column of the
Young diagram characterizing the corresponding irrep.
We will focus on  theories describing gauge fields
$\phi_{\m_1 \dots \m_p \vert\n_1 \dots \n_q}$ whose symmetries
correspond to the Young diagram $[p,q]$ formed by two columns of
arbitrary (but different) lengths $p$ and $q$ ($p>q\,$).
The physical degrees of freedom for such theories
correspond to a traceless tensor carrying an irrep. of the little
group $O(D-2)$ associated with the Young diagram $[p,q]\,$.
Therefore, we will work in spacetime dimension $D\geqslant p+q+2$
so that the field carries local physical degrees of freedom.
Such fields were studied recently at the free level in AdS background
\cite{Alkalaev:2003hc,deMedeiros:2003px}.
In the sequel, we will frequently use a loose terminology by
referring to a tensor irrep. by  its Young diagram.
\vspace{.4cm}

In the present paper, we address the natural  problem of switching
on consistent self-interactions among $[p,q]$-type tensor gauge
fields in flat background, where $p\neq q\,$. As in
\cite{Barnich:1993pa,Henneaux:1997ha,Boulanger:2000rq,Boulanger:2004rx,Bekaert:2002uh,Bizdadea:2003ht},
we use the BRST-cohomological reformulation of the Noether method
for the problem of consistent interactions \cite{Barnich:vg}. For
an alternative Hamiltonian-based deformation point of view, see
\cite{Bizdadea:2001re}. The question of consistent
self-interactions in flat background has already been investigated
in the case of vector (\ie $[1,0]$) gauge fields in
\cite{Barnich:1993pa}, $p$-forms (\ie $[p,0]$-fields) in
\cite{Henneaux:1997ha}, Fierz-Pauli $[1,1]$-fields in
\cite{Boulanger:2000rq}, $[p,1]$-fields ($p>1\,$) in
\cite{Bekaert:2002uh}, $[2,2]$-fields in \cite{Bizdadea:2003ht}
and $[p,p]$-fields ($p>1\,$) in \cite{Boulanger:2004rx}. Here, we
extend and strengthen the results of \cite{Bekaert:2002uh} by
relaxing some assumptions on the number of derivatives in the
interactions. The present work is thus the completion of the
analysis of self-interactions for \textit{arbitrary} $[p,q]$-type
tensor gauge fields in flat space. \vspace{.4cm}

Our main (no-go) result can be stated as follows, spelling out explicitly
our assumptions:
\newpage
\noindent{\bfseries Theorem:}
{\textit{In flat space and under the assumptions of locality and
translation-invariance, there is no consistent smooth deformation of the
free theory for $[p,q]$-type tensor gauge fields with $p\neq q$ that
modifies the gauge algebra.
Furthermore, for $q>1$, when there is no positive integer $n$ such that
$p+2=(n+1)(q+1)$, there exists no smooth deformation that alters the
gauge transformations either. Finally, if one excludes deformations that
involve four derivatives or more in the Lagrangian and that 
are not Lorentz-invariant, then there is no
smooth deformation at all.}}

\vspace*{.3cm}

The paper is organized as follows. In Section \ref{freetheory}, we
review the free theory of $[p,q]$-type tensor gauge fields. In
Section \ref{BRST}, we introduce the BRST construction for the
theory. Sections \ref{cohogamma} to
\ref{Invariantcharacteristiccohomology} are devoted to the proof
of cohomological results. We compute $H(\g)$ in Section
\ref{cohogamma}, an invariant Poincar\'e lemma is proved in
Section \ref{InvariantPoincarelemma}, the  cohomologies $H_k^D(\d
\vert d)$ and $H_k^{D\, inv}(\d \vert d)$ are computed
respectively in Sections \ref{Characteristiccohomology} and
\ref{Invariantcharacteristiccohomology}. The self-interaction
question is answered in Section \ref{self-interactions}. A
brief concluding section is finally followed by three appendices
containing the proofs of three theorems presented in the core of
the paper.

\section{Free theory}
\label{freetheory}

As stated above, we consider theories for mixed tensor gauge
fields $\phi_{\m_1 \dots \m_{p} \vert \n_1 \dots \n_q }$ whose
symmetry properties are characterized by two columns of arbitrary
(but different) lengths. In other words, the gauge field obeys  the
conditions
\bqn
&\phi_{\m_1 \dots \m_{p} \vert \n_1 \dots \n_q }
=\phi_{[\m_1 \dots \m_{p}] \vert \n_1 \dots \n_q
}=\phi_{\m_1 \dots \m_{p} \vert [\n_1 \dots \n_q]}\,,&
\nonumber \\
&\phi_{[\m_1 \dots \m_{p} \vert \n_1]\n_2 \dots \n_q }=0\,,&
\nonumber
\eqn
where square brackets denote strength-one complete antisymmetrisation.

\subsection{Lagrangian and gauge invariances}
\label{Lagrangianandgaugeinvariance}

The Lagrangian of the free theory is \bqn \cl =
-\frac{1}{2\,(p+1)!\,q!}\;\delta^{[\r_1 \dots \r_q \m_1 \dots
\m_{p+1}]}_{[\n_1 \dots \n_q \s_1 \dots \s_{p+1}]}\;
\pa^{[\s_1}\phi^{\s_2 \dots \s_{p+1}]\vert}_{\hspace*{1.4cm} \r_1
\dots \r_q}\; \pa^{}_{[\m_1} \phi_{\m_2 \dots
\m_{p+1}]\vert}^{\hspace*{1.4cm}\n_1 \dots \n_q }\,, \nn \eqn
where the generalized Kronecker delta has strength one. This
Lagrangian was obtained for $[2,1]$-fields in
\cite{Curtright:1980yk}, for $[p,1]$-fields in \cite{Aulakh:cb}
and, for the general case of $[p,q]$-fields, in the second paper
of \cite{deMedeiros:2002ge}.

The quadratic action \bqn S_0 [\phi]=\int d^Dx\,{\cal
L}(\partial\phi) \label{action} \eqn is invariant under gauge
transformations with gauge parameters $\a^{(1,0)}$ and
$\a^{(0,1)}$ that have respective symmetries $[p-1,q]$ and
$[p,q-1]\,$. In the same manner as $p$-forms, these gauge
transformations are \textit{reducible}, their order of
reducibility growing with $p$. We identify the gauge field
$\phi\,$ with $\a^{(0,0)}$, the zeroth order parameter of
reducibility. The gauge transformations and their reducibilities
 are\footnote{We
introduce the short notation $\m_{[p]}\equiv [\m_1 \dots \m_{p}]\,$.
A comma stands for a derivative: $\a_{,\n}\equiv \pa_{\n}\a$.}
\bqn \d \a^{(i,j)}_{\m_{[p-i]} \vert \n_{[q-j]}}&=
&\pa_{[\m_1}\a^{(i+1,j)}_{\m_2 \dots \m_{p-i}]
\vert\, \n_{[q-j]}}\nonumber \\
 &&+\, b_{i,j}  \left( \a^{(i,j+1)}_{\m_{[p-i]} \vert\, [\n_{[q-j-1]},\n_{q-j}]}
+ a_{i,j}\,\a^{(i,j+1)}_{\n_{[q-j]}[\m_{q-j+1} \dots
\m_{p-i}\vert\, \m_{[q-j-1]},\m_{q-j}]} \right)
\label{gaugetransfo} \eqn where $i=0, ..., p-q$ and $j=0, ...,
q\,$. The coefficients $a_{i,j}$ and $b_{i,j}$ are given by \bqn
a_{i,j}=\frac{(p-i)!}{(p-i-q+j+1)! \, (q-j)!}\,,\quad
b_{i,j}=(-)^i \, \frac{(p-q+j+2)}{(p-i-q+j+2)}\,.\nonumber \eqn To
the above formulae, we must add the convention that, for all
$j\,$,  $\a^{(p-q+1,j)}=0=\a^{(j,q+1)}\,$. The symmetry properties
of the parameters $\a^{(i,j)}$ are those of Young diagrams with
two columns of lengths $p-i$ and $q-j\,$. More details on the
reducibility parameters $\a^{(i,j)}_{\m_1 \dots \m_{p-i}\vert\,
\n_1 \dots \n_{q-j}}$ will be given in Subsection
\ref{BRSTghostsofghosts}.

The fundamental gauge-invariant object is the field strength
$K\,$, the $[p+1,q+1]$-tensor defined as the double curl of the
gauge field \bqn K_{\m_1 \dots \m_{p+1} \vert\, \n_1 \dots
\n_{q+1} }\equiv\pa_{[\m_1}\phi_{\m_2 \dots \m_{p+1} ]\,\vert\, [\n_1
\dots \n_q \,,\,\n_{q+1} ]}\,.\nonumber \eqn By definition, it
satisfies the Bianchi (BII) identities \be
\partial_{[\m_1} K_{\m_2 \dots \m_{p+2}] \vert\, \n_1 \dots \n_{q+1} }=0\,,
\quad K_{\m_1 \dots \m_{p+1} \,\vert\, [\n_1 \dots \n_{q+1},\n_{q+2}]}=0\,.
\label{Bianchi}
\ee
The field strength tensor $K$ plays a crucial role in the
determination of the physical degrees of freedom described
by the action $S_0[\phi]\,$.

\subsection{Equations of motion}
\label{equationsofmotion}

The equations of motion are expressed in terms of the field
strength: \bqn G^{\m_1 \dots \m_p \vert\,}_{ \hspace*{1cm}\n_1
\dots \n_q} \equiv \frac{\d {\cal L}}{\d\phi_{\m_1 \dots \m_p
\vert\,}^{\hspace*{1cm} \n_1 \dots \n_q}}=
\frac{1}{(p+1)!q!}\;\delta^{[\r_1 \dots \r_{q+1} \m_1 \dots
\m_{p}]}_{[\n_1 \dots \n_q \s_1 \dots \s_{p+1}]}\; K^{\s_1\dots
\s_{p+1}\vert\,}_{\hspace*{1.3cm}\r_1 \dots \r_{q+1}}\approx 0 \,,
\nonumber \eqn where a weak equality ``$\approx$'' means ``equal
on the surface of the solutions of the equations of motion''.
This is a generalization of vacuum Einstein equations, linearized
around the flat background. Taking successive traces of the
equations of motion, one can show that they are equivalent to the
tracelessness of the field strength \be \eta^{\s_1\r_1}K_{\s_1
\dots \s_{p+1}\vert\, \r_1 \dots \r_{q+1}}\approx 0\,.
\label{Ricci} \ee This equation generalizes the vanishing of the
Ricci tensor (in the vacuum), and is non-trivial only when $p+q+2
\leqslant D$. Together with the ``Ricci equation" (\ref{Ricci}),
the Bianchi identities (\ref{Bianchi}) imply \cite{Hull:2001iu}
\be
\partial^{\s_1}K_{\s_1 \dots \s_{p+1}\vert\, \r_1 \dots \r_{q+1}}\approx 0
\approx \partial^{\r_1}K_{\s_1 \dots \s_{p+1}\vert\, \r_1\dots \r_{q+1}}\,.
\label{divergence}
\ee
The gauge invariance of the action is equivalent to the
divergenceless of the tensor $G^{\m_{[p]} \vert \n_{[q]}}$, that is, the latter
satisfies the Noether identities
\be
\partial^{\s_1}G_{\s_1 \dots \s_{p+1}\vert\, \r_1 \dots
\r_{q+1}}=0=\partial^{\r_1}G_{\s_1 \dots \s_{p+1}\vert\, \r_1
\dots \r_{q+1}}\,. \label{Noether} \ee These identities are a
direct consequence of the Bianchi ones (\ref{Bianchi}). The
Noether identities (\ref{Noether}) ensure that the equations of
motion can be written as \bqn 0\approx G^{\m_1 \dots \m_p \vert\,
\n_1 \dots \n_q} =\pa_{\a}H^{\a\m_1 \dots \m_p \vert\, \n_1 \dots
\n_q}\,, \nonumber \eqn where  \bqn H^{\a\m_1 \dots \m_p \vert\,}_{ \hspace*{1.2cm}\n_1 \dots \n_q}
=\frac{1}{(p+1)!q!} \;\delta^{[\r_1 \dots \r_q \a \m_1 \dots
\m_{p}]}_{[\n_1 \dots \n_q \b \s_1 \dots \s_p]}\;
\partial^{[\beta}\phi^{\s_1 \dots \s_p]\vert\,}_{\hspace*{1.1cm}\r_1
\dots \r_q} \,. \nonumber \eqn The symmetries of the tensor
$H$ correspond to the Young diagram $[p+1,q]\,$. This property will be useful in the
computation of the local BRST cohomology.

\subsection{Physical degrees of freedom}

The ``Ricci equation" (\ref{Ricci}) states that, on-shell, the
field strength belongs to the irrep. $[p+1,q+1]$ of $O(D-1,1)\,$.
The Bianchi identities together with (\ref{divergence}) further
imply that the on-shell non-vanishing components of the field
strength belong to the unitary irrep. $[p,q]$ of the little group
$O(D-2)$. Indeed, on-shell, gauge fields in the light-cone gauge
are essentially field strengths \cite{Siegel:1986zi}, and
the ``Ricci equation" takes the form $$ \d^{i_1j_1}\phi_{i_1
\dots i_{p}\vert\, j_1 \dots j_{q}}\approx 0\,. $$ where $i$ and
$j$ denote light-cone indices ($i,j=1,\ldots,D-2\,$). As a
consistency check, one can note that the latter equation is
non-trivial only when $p+q\geqslant D-2$. The theory describes the
correct physical degrees of freedom of a first-quantized
massless particle propagating in flat space, \ie, the latter particle
provides a unitary irrep. of the group $IO(D-1,1)\,$.

We should stress that the exact analogue of all the previous
properties hold for arbitrary mixed symmetry fields. This result
was obtained by two of us and was mentioned in
\cite{Bekaert:2003zq} but the detailed proof was not given
there\footnote{The proof presented in this paper (Appendix A) provides an
indirect proof that the light-cone gauge is reachable (so that the
theory describes the correct number of physical degrees of freedom). We
would like to underline the fact that the works
\cite{Hull:2001iu,deMedeiros:2002ge} assume (but do not contain
any rigorous proof of) this fact. It would not be straightforward
to prove it directly because the tower of ghosts is extremely
complicated in the general case.}. We take the opportunity to
provide this extremely simple proof in Appendix
\ref{BargmannWigner} for the particular case of two-column gauge
fields, since it already covers all the features of the general
case for arbitrary mixed tensor gauge fields.  

\section{BRST construction}
\label{BRST}

\subsection{BRST deformation technique}
\label{BRSTdefo}

Once one has a consistent free theory, it is natural to try to
deform it into an interacting theory. The traditional Noether
deformation procedure assumes that the deformed action can be
expressed as a power series in a coupling constant $g\,$, the
zeroth-order term in the expansion describing the free theory $S_0\,$.
The procedure is perturbative: one tries to construct the deformations
order by order in the deformation parameter $g\,$.

Some physical requirements naturally come out:
\begin{itemize}
  \item \underline{non-triviality}: we reject {\em trivial} deformations
arising from field-redefinitions that reduce to the identity at
order $g^0\,$: \bqn \phi\longrightarrow \phi'=\phi+g\, \theta (\phi, \pa
\phi, \cdots)+\co(g^2)\,. \label{fieldredef} \eqn
  \item \underline{consistency}: a deformation of a
theory is called {\em{consistent}} if the deformed theory
possesses the same number of (possibly deformed) independent
gauge symmetries, reducibility identities, {\it etc.}, as the system we
started with. In other words, the number of physical degrees of
freedom is unchanged.
  \item \underline{locality}: The deformed action $S[\phi]$ must be a {\em local}
  functional. The deformation of the gauge transformations, {\it etc.},
  must be local functions, as well as the field redefinitions.
\end{itemize}
We remind the reader that a local function of some set of fields $\varphi^i$ is a smooth function
of the fields $\varphi^i$ and their
derivatives $\partial\varphi^i$, $\partial^2\varphi^i$, ... up to some
{\it finite} order, say $k$, in the number of derivatives.
Such a set of variables $\varphi^i$, $\partial\varphi^i$, ...,
$\partial^k\varphi^i$ will be collectively denoted by $[\varphi^i]$.
Therefore, a local function of $\varphi^i$ is denoted by $f([\varphi^i])$. A local
$p$-form $(0\leqslant p \leqslant D)$ is a differential $p$-form the components of which are local
functions: 
\bqn \omega =
\frac{1}{p!}\,\omega_{\m_1\ldots\m_p}(x, [\phi^i])\,
dx^{\m_1} \wedge \cdots \wedge dx^{\m_{p}}\,.\nonumber 
\eqn
A local functional is the integral of a local $D$-form.

As shown in \cite{Barnich:vg}, the Noether procedure can be
reformulated in a BRST-cohomological formalism: the
first-order non-trivial consistent local interactions are in
one-to-one correspondence with elements of the cohomology
$H^{D,0}(s \vert\, d)$ of the BRST differential $s$ modulo the
total derivative $d\,$, in maximum form-degree $D$ and in ghost
number $0\,$. That is, one must compute the general solution of
the cocycle condition \be s a^{D,0} + db^{D-1,1} =0, \label{coc}
\ee where $a^{D,0}$ is a top-form of ghost number zero and
$b^{D-1,1}$ a $(D-1)$-form of ghost number one, with the
understanding that two solutions of (\ref{coc}) that differ by a
trivial solution should be identified \bqn a^{D,0}\sim a^{D,0} + s
m^{D,-1}  + dn^{D-1,0} \nonumber \eqn as they define the same
interactions up to field redefinitions (\ref{fieldredef}). The
cocycles and coboundaries $a,b,m,n,\ldots\,$ are local forms of
the field variables (including ghosts and antifields)

\subsection{BRST spectrum}\label{BRSTghostsofghosts}
\label{BRSTspectrum}

In the theories under consideration and according to the general
rules of the BRST-antifield formalism, one associates with each
gauge parameter $\a^{(i,j)}$ a ghost, and then to any field
(including ghosts) a corresponding antifield (or antighost) of opposite Grassmann
parity. More precisely, the spectrum of fields (including ghosts)
and antifields is given by
\begin{itemize}
\item {\underline{the fields}}: $A^{(i,j)}_{\m_{[p-i]}
\vert\,\n_{[q-j]}}\,$, where $A^{(0,0)}$ is identified with
$\phi\,$; \item {\underline{the antifields}}:
$A^{*(i,j)\;\m_{[p-i]}\vert\, \n_{[q-j]}}\,$,
\end{itemize}
where  $i=0, ..., p-q$ and $j=0,..., q\,$. The symmetry properties
of the fields $A^{(i,j)}_{\m_{[p-i]} \vert\,\n_{[q-j]}}$ and
antifields $A^{*(i,j)\;\m_{[p-i]}\vert\, \n_{[q-j]}}$ are those of
Young diagrams with two columns of lengths $p-i$ and $q-j\,$. To
each field and antifield are associated a pureghost number and an
antifield (or antighost) number. The pureghost number is given by
$i+j$ for the fields $A^{(i,j)}$ and $0$ for the antifields, while
the antifield number is $0$ for the fields and $i+j+1$ for the
antifields $A^{*(i,j)}\,$. The Grassmann parity is given by the
pureghost number (or the antighost number) modulo $2\,$. All
this is summarized in Table \ref{table1}.

\begin{figure}[ht!]
\hspace*{.6cm}
\xymatrix @!=.3cm {
 && & & & & [p,q]\ar@/l3cm/[ddddddllllll]^{(p-q)\,steps}\ar[dr]\ar[dl] &  &  & & &\\
 && & & & [p-1,q]\ar[dr]\ar[dl] &  & [p,q-1] \ar[dr]\ar[dl] &  && &\\
 && & & [p-2,q]\ar@{-->}[ddddllll] &  & [p-1,q-1] &  & [p,q-2]\ar@{-->}[ddrr] && & \\
 && & & &\ldots &   &  \ldots&  &  & &  &  \\
 &&&~~\ldots& &\ldots\ldots & & &\ldots\ldots & & [p\,,0]~(2q< p) \ar@{-->}[ddddddllll]&& \\
 &&&\ldots&&\txt{\footnotesize{$[q+i,q\!-\!i\!+\!1]$}}\ar[dr]\ar[dl]&&\ldots&&&&&  \\
[q,q]\ar@/l3cm/[ddddrrrrrr]^{q\,\,steps}\ar@{-->}[ddddrrrrrr]&&\ldots\ldots &
  &\txt{\footnotesize{$[q\!+\!i\!-\!1,q\!-\!i\!+\!1]$}}&& ~~\txt{\footnotesize{$[q+i,q-i]$}}
  &&\ldots\ldots&~\;\ldots&&\ar@{-->}[ddr][p,2q-p]~(2q \geqslant p) &\\
 && &  \ldots& &  & \ldots\ldots &  &  &  & \ldots &  & \\
 && &  &  & \ldots\ldots &  & &&\ldots\ldots &  &  & [p\,,0]\ar@{-->}[ddllllll]\\
 && &  &  &  &  &  &  &  &  &  & \\
 && &  &  &  & [q\,,0] &  &  &  &  &  & \\
}
\caption{\textit{Antighost-zero BRST spectrum of $[p,q]-$type gauge field. \label{figure1}}}
\end{figure}
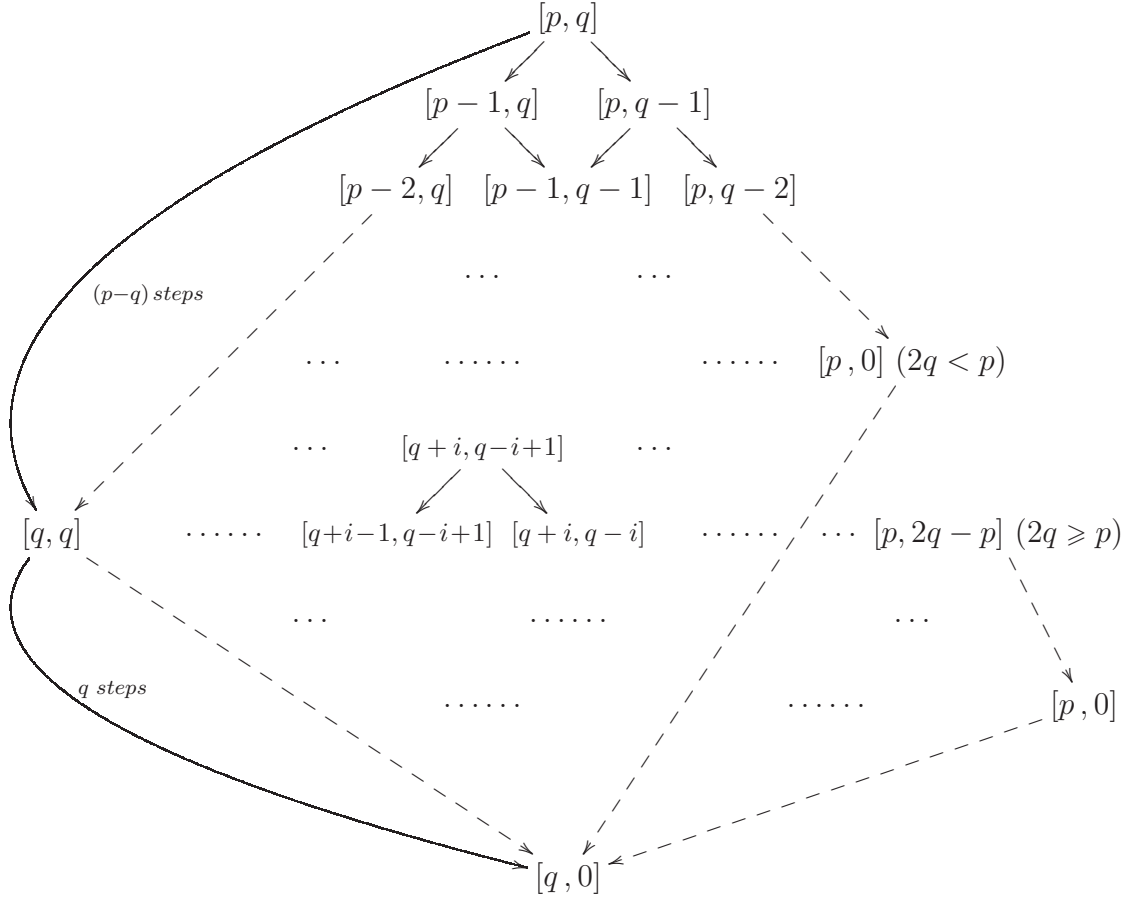

\begin{table}[!ht]
  \centering
\begin{tabular}{|c|c|c|c|c|}\hline
  & Young & $puregh$ & $antigh$ & Parity \\\hline
  $A^{(i,j)}$ & $[p-i,q-j]$ & $i+j$ & $0$ & $i+j$ \\\hline
  $A^{*(i,j)}$ & $[p-i,q-j]$ & $0$ & $i+j+1$ & $i+j+1$ \\ \hline
\end{tabular}
\caption{\it Symmetry, pureghost number, antighost number
and parity of the (anti)fields.\label{table1}}
\end{table}
One can visualize the whole BRST spectrum in vanishing antighost
number as well as the procedure that gives all the ghosts
starting from $\phi_{\m_{[p]}\,\vert\, \n_{[q]}}$ on Figure \ref{figure1}, where the pureghost number increases from top down, by
one unit at each line.
At the top of Figure \ref{figure1} lies the gauge field
$\phi_{\m_{[p]}\,\vert\, \n_{[q]}}$ with pureghost number zero. At
the level below, one finds the pureghost number one gauge
parameters $A^{(1,0)}_{\m_{[p-1]}\vert\, \n_{[q]}}$ and
$A^{(0,1)}_{\m_{[p]}\vert\, \n_{[q-1]}}\,$ whose respective
symmetries are obtained by removing a box in the first (resp.
second) column of the Young diagram $[p,q]$ corresponding to the
gauge field $\phi_{\m_{[p]}\,\vert\, \n_{[q]}}\,$ (the rules that
give the $(i+1)$-th generation ghosts from the $i$-th generation
ones can be found in \cite{Labastida:1986gy,Bekaert:2002dt}).
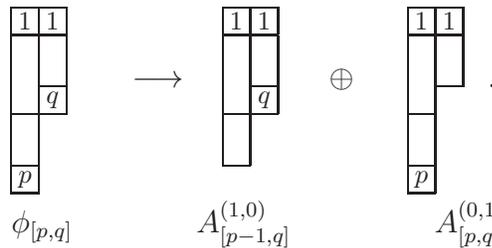
\vspace*{1cm}

\begin{center}
\begin{picture}(150,50)(0,-60)
\multiframe(0,10)(10.5,0){1}(10,10){\ft$ 1$}
\multiframe(0,-20)(10.5,0){1}(10,29.5){$ $}
\multiframe(10.5,10)(10.5,0){1}(10,10){\ft$1 $}
\multiframe(10.5,-9.5)(10.5,0){1}(10,19){$ $}
\multiframe(10.5,-20)(10.5,0){1}(10,10){\ft$q$}
\multiframe(0,-39.5)(10.5,0){1}(10,19){$ $}
\multiframe(0,-50)(10.5,0){1}(10,10){\ft$p$}
\put(30,-10){~~~~$\longrightarrow$}
\multiframe(80,10)(10.5,0){1}(10,10){\ft$ 1$}
\multiframe(80,-20)(10.5,0){1}(10,29.5){$ $}
\multiframe(90.5,10)(10.5,0){1}(10,10){\ft$1 $}
\multiframe(90.5,-9.5)(10.5,0){1}(10,19){$ $}
\multiframe(90.5,-20)(10.5,0){1}(10,10){\ft$q$}
\multiframe(80,-39.5)(10.5,0){1}(10,19){$ $}
\put(120,-10){{$\oplus$}}
\multiframe(150,10)(10.5,0){1}(10,10){\ft$ 1$}
\multiframe(150,-20)(10.5,0){1}(10,29.5){$ $}
\multiframe(160.5,10)(10.5,0){1}(10,10){\ft$1 $}
\multiframe(160.5,-9.5)(10.5,0){1}(10,19){$ $}
\multiframe(150,-39.5)(10.5,0){1}(10,19){$ $}
\multiframe(150,-50)(10.5,0){1}(10,10){\ft$p$}
\put(180,-10){.}\put(0,-65){$\phi_{[p,q]}$}
\put(70,-65){$A^{(1,0)}_{[p-1,q]}$}
\put(160,-65){$A^{(0,1)}_{[p,q-1]}$}
\end{picture}
\end{center}

In pureghost number $p-q$, we obtain a set of ghosts containing
$A^{(p-q,0)}_{\m_{[q]}\n_{[q]}}$ $\sim [q,q]\,$. The Young diagram
corresponding to the latter ghost is obtained by removing $p-q$
boxes from the first column of $[p,q]$.
\vspace{.2cm}

 If $q<p-q\,$, we do not have to reach the pureghost
level $p-q$ to find the $p$-form ghost $A^{(0,q)}_{\m_{[p]}}\,$
$\sim~[p\,,0]\,$. If $2q \geqslant p\,$, we have to remove
additional boxes from the second column of $[p,q]$ in order to
empty it completely and obtain the $p$-form ghost
$A^{(0,q)}_{\m_{[p]}}$. The Young diagrams of the remaining ghosts
are obtained by further removing boxes from the Young diagram
corresponding to the ghost $A^{(0,q)}_{\m_{[p]}}$ with
$puregh=q\,$. This procedure will terminate at pureghost number
$p$ with the $q$-form ghost $A^{(p-q,q)}_{\m_{[q]}}\sim [q,0]\,$.
It is not possible to find ghosts $A_{\m_{[r]}\vert \n_{[s]}}$
with $r,s<q\,$, since it would mean that two boxes from a same row
would have been removed from $[p,q]$, which is not allowed
\cite{Labastida:1986gy,Bekaert:2002dt}.

The antighost sector has exactly the same structure as the ghost sector of
Figure \ref{figure1}, where each ghost $A^{(i,j)}$ is replaced by its antighost $A^{*(i,j)}$.

\subsection{BRST differential}
\label{BRSTdifferential}

The BRST differential $s$ of  the free theory (\ref{action}), (\ref{gaugetransfo}) is generated by the functional
\bqn
W_0 = S_0 [\phi] \;+ \int d^Dx\; &\Big[& \sum_{i=0}^{p-q} \sum_{j=0}^{q}
(-)^{i+j} \,
A^{*(i,j)\;\m_1 \dots \m_{p-i} \vert\, \n_1 \dots \n_{q-j}} \nonumber \\
&\,&\times(\pa^{}_{[\m_1}A^{(i+1,j)}_{\m_2 \dots \m_{p-i}] \vert\,
\n_1 \dots \n_{q-j}} - b_{i+1,j}\,A^{(i,j+1)}_{\m_1 \dots \m_{p-i}
\vert\, [\n_1 \dots \n_{q-j-1},\n_{q-j}]})\Big]\;, \nonumber \eqn
with the convention that $ A^{(p-q+1,j)}= A^{(i,q+1)}=
A^{*(-1,j)}= A^{*(i,-1)}=0\,$. More precisely, $W_0$ is the
generator of the BRST differential $s$ of the free theory through
\bqn s A = (W_0, A)_{a.b.}\,, \nonumber \eqn where the antibracket
$(~,~)_{a.b.}$ is defined by $(A,B)_{a.b.}=\frac{\d^R A}{\d
\Phi^I}\frac{\d^L B}{\d \Phi^*_I}- \frac{\d^R A}{\d
\Phi^*_I}\frac{\d^L B}{\d \Phi^I}\,$. The functional $W_0$ is a
solution of the \emph{master equation}
$$(W_0,W_0)_{a.b.}=0\,.$$
The BRST-differential
$s$ decomposes into $s=\g + \d \,$. The first piece $\g\,$, the differential along the gauge orbits,
increases the pureghost number by one unit, whereas the Koszul-Tate differential $\d$
decreases the antighost (or antifield) number by one unit.
A $\mathbb{Z}$-grading called \emph{ghost number} (or $gh$) corresponds to the
differential $s\,$. We have $$gh=puregh - antigh \,.$$
The action of $\g$ and $\d$ on the BRST variables is zero, except
\bqn
\g A^{(i,j)}_{\m_{[p-i]} \vert\, \n_{[q-j]}}
&= &\pa_{[\m_1}A^{(i+1,j)}_{\m_2 \dots \m_{p-i}]
\vert\,
\n_{[q-j]}}\nonumber \\
 &&+\, b_{i,j}  \left( A^{(i,j+1)}_{ \m_{[p-i]} \vert\,
 [\n_{[q-j-1]},\n_{q-j}]} +
 a_{i,j}A^{(i,j+1)}_{\n_{[q-j]}[\m_{q-j+1} \dots \m_{p-i}\vert\,
\m_{[q-j-1]},\m_{q-j}]}\right)\nonumber\\
\d  A^{*(0,0)\; \m_{[p]} \vert\,  \n_{[q]} } &=& G^{\m_{[p]}
\vert\,\n_{[q]}}\nonumber \\
\d  A^{*(i,j)\;\m_{[p-i]} \vert\,  \n_{[q-j]} }
&=&
(-)^{i+j}\left( \pa_{\s}A^{*(i-1,j)\;\s\m_{[p-i]} \vert\,
\n_{[q-j]}} -\frac{1}{p-i+1}\,\pa_{\s}A^{*(i-1,j)\; \n_1\m_{[p-i]}
\vert\, \s \n_2 \dots \n_{q-j}} \right)
\nonumber \\
&&+(-)^{i+j+1}b_{i+1,j-1}\pa_{\s} A^{*(i,j-1)\;\m_{[p-i]} \vert\,\n_{[q-j]}\s}\,,
\nonumber
\eqn
where the last equation holds only for $(i,j)$ different from $(0,0)$.

For later computations, it is useful to define a unique antifield
for each antighost number: \bqn C^{* \; \m_1 \dots \m_q \vert\,
\n_1 \dots \n_j}_{p+1-j}=
\sum_{k=0}^{j}\epsilon_{k,j}A^{*(p-q-j+k,q-k)\;\m_1 \dots \m_q
[\n_{k+1}\dots \n_j \vert\, \n_1 \dots \n_k]} \nonumber \eqn for
$0 \leqslant j \leqslant p\,$, and, in antighost zero, the
following specific combination of single derivatives of the field
\bqn C^{* \; \m_1 \dots \m_q \vert\, \n_1 \dots \n_{p+1}}_{0}=
\epsilon_{q,p+1}H^{\m_1 \dots \m_q [\n_{q+1}\dots \n_{p+1} \vert\,
\n_1 \dots \n_q]} \,, \nonumber \eqn where $\epsilon_{k,j}$
vanishes for $k>q$  and for $j-k>p-q\,$, and is given in the other
cases by: \bqn \epsilon_{k,j}=(-)^{pk+j(k+p+q)+
\frac{k(k+1)}{2}}\frac{(^k_{p+1})\, (^k_j)}{(^k_q)} \nonumber \eqn
where $(_n^m)$ are the binomial coefficients ($n\geqslant m$). Some properties of
the new variables $C^{*}_{k}$ are summarized in Table
\ref{table2}.

\begin{table}[!ht]
  \centering
\begin{tabular}{|c|c|c|c|c|}\hline
  & Young diagram & $puregh$ & $antigh$ & Parity \\\hline
  $C^{*}_{k}$ & $[q] \otimes [p+1-k]- [p+1] \otimes [q-k] $ & $0$ & $k$ & $k $\\ \hline
\end{tabular}
\caption{\it Young representation, pureghost number, antighost
number and parity of the antifields $C^{*}_{k}$.}\label{table2}
\end{table}
\noindent The symmetry properties of $C^{*}_{k}$ are denoted by
\bqn [q] \otimes [p+1-k]\,-\,[p+1] \otimes [q-k] \nonumber \eqn
which means that they have the symmetry properties corresponding
to the tensor product of a column $[q]$ by a column $[p+1-k]$ from
which one should substract (when $k\leqslant q$) all the Young
diagrams appearing in the tensor product $[p+1] \otimes [q-k]$.
\vspace*{.2cm}

The antifields $C^{*\; \m_{[q]} \vert\, \nu_{[p+1-k]}}_k$ have
been defined in order to obey the following relations: \bqn \d
C^{* \; \m_1 \dots \m_q \vert\, \n_1 \dots \n_j}_{p+1-j} &=&
\pa_{\s} C^{* \;\m_1 \dots \m_q \vert\, \vert\, \n_1 \dots \n_j
\s}_{p-j}\quad
{\rm{for}} \quad 0 \leqslant j \leqslant p \,,\nonumber \\
\d C^{* \; \m_1 \dots \m_q \vert\, \n_1 \dots \n_{p+1}}_{0}
&=&0\,. \label{beau} \eqn
 If we further define the
inhomogeneous form \bqn \tilde{H}^{\m_1 \dots \m_q
}\equiv\sum_{j=0}^{p+1}C^{* \, D-j\; \m_1 \dots \m_q  }_{p+1-j}\;,
\nonumber \eqn where  \bqn C^{* \, D-j\; \m_1 \dots \m_q
}_{p+1-j}\equiv (-)^{jp+\frac{j(j+1)}{2}}\frac{1}{j!(D-j)!} \,C^{*
\; \m_1 \dots \m_q \vert\,\n_1 \dots \n_j}_{p+1-j}\epsilon_{\n_1
\ldots\n_D}dx^{\n_{j+1}}\ldots dx^{\n_D}\,, \nonumber \eqn then, as a
consequence of (\ref{beau}), any polynomial $P(\tilde{H})$ in
$\tilde{H}^{\m_1 \dots \m_q }$ will satisfy \bqn
(\d+d)P(\tilde{H})=0\,. \label{htilde} \eqn

The polynomial $\tilde{H}$ is not invariant under gauge transformations. It is therefore useful to introduce another polynomial, $\tilde{\cal H}\,$, with an explicit $x$-dependance, that {\it is} invariant. $\tilde{\cal H}\,$ is defined by $$\tilde{\cal H}_{\m_{[q]}}\equiv \sum_{j=1}^{p+1}C^{* \, D-p-1+j}_{j\, \m_{[q]}}+\tilde{a} \,\epsilon_{[\m_{[q]}\s_{[p+1]} \t_{[D-p-q-1]}]} K^{q+1\,\s_{[p+1]}}x^{\t_1}dx^{\t_2} \ldots dx^{\t_{D-p-q-1}}\,,$$
where $\tilde{a} =(-)^{\frac{p(p-1)+q(q-1)}{2}}\frac{1}{q!q!(p+q+1)!(p+1-q)!(D-p-q-1)!}$.  One can check that $\tilde{\cal H}=\tilde{H}+dm^{D-p-2}_0$. This fact has the consequence that polynomials in $\tilde{\cal H}$ also satisfy $(\d+d)P(\tilde{\cal H})=0$.

\section{Cohomology of $\g$}
\label{cohogamma}

We hereafter give the content of $H(\g)$. Subsequently, we
explain the procedure that we followed in order to obtain that
result.

\begin{theorem}\label{Hgamma} The cohomology of $\g$ is isomorphic to the space of
functions depending on
\begin{itemize}
  \item the antifields and their derivatives $[A^{*(i,j)}]\,$,
  \item the curvature and its derivatives $[K]\,$,
  \item the $p\,$-th generation ghost $A^{(p-q,q)}$ and
  \item the curl $D^0_{\m_1 \ldots \m_{p+1}} \equiv (-)^q\pa^{}_{[\m_1}A^{(0,q)}_{\m_2 \ldots
\m_{p+1} ]}$ of the $q\,$-th generation ghost $A^{(0,q)}$.
\end{itemize}
\begin{eqnarray}
    H(\g)\simeq \left\{ f\left([A^{*(i,j)}],[K],A^{(p-q,q)},D^0_{\m_1 \ldots \m_{p+1}}
     \right)\right\}\nn\,.
\end{eqnarray}
\end{theorem}

\vspace*{.2cm} \noindent{ \bfseries{Proof :}} The antifields and
all their derivatives are annihilated by $\g\,$. Since they carry
no pureghost degree by definition, they cannot be equal to the
$\g\,$-variation of any quantity. Hence, they obviously belong
to the cohomology of $\g\,$.

To compute the $\g\,$-cohomology in the sector of the field, the
ghosts and all their derivatives, we split the variables into
three sets of {\it independent} variables obeying respectively $\g
u^{\ell} = v^{\ell}\,$, $\g v^{\ell} = 0\,$ and $\g w^i=0\,$. The
variables $u^{\ell}$ and $v^{\ell}$ form so-called ``contractible
pairs" and the cohomology of $\g$ is therefore generated by the
variables $w^i\,$ (see e.g. \cite{book}, Theorem 8.2).

We decompose the spaces spanned by the derivatives
$\pa_{\m_1\ldots\m_k}A^{(i,j)}\,$, $k\geqslant 0\,$, $0\leqslant i
\leqslant p-q\,$, $0\leqslant j\leqslant q\,$, into irreps of
$GL(D,\mathbb{R})\,$ and use the structure of the reducibility
conditions (see Figures  2. and 3.) in order to group the
variables into contractible pairs.

\hspace*{2cm}
\xymatrix @!=.3cm {
 & & \\
 & A^{(i,j)}\ar@{~}[dl]_{d^{\{1\}}}\ar@{~}[dr]^{d^{\{2\}}}\ar@{~}[ul]_{d_{\{2\}}}& \\
  A^{(i+1,j)} & & A^{(i,j+1)} \\}
  \hspace*{3cm}
\xymatrix @!=.3cm {
 A^{(i,j-1)}\ar@{~}[dr]_{d^{\{2\}}} & & A^{(i-1,j)}\ar@{~}[dl]^{d^{\{1\}}} \\
  & A^{(i,j)}\ar@{~}[dr]_{d^{\{2\}}}&  \\
  & & \\}

\vspace*{.1cm}
\hspace*{3.5cm}{\footnotesize{Figure 2}}\hspace*{5.6cm}{\footnotesize{Figure 3}}
\vspace*{.2cm}

\noindent We use the differential operators $d^{\{i\}}\,$,
$i=1,2,...$ (see \cite{Bekaert:2002dt} for a general definition)
which act, for instance on Young-symmetry type tensor fields
$T_{[2,1]}$, as follows:

\vspace*{.5cm}\hspace*{1.5cm}
\begin{picture}(38,45)(-10,-30)
\put(-30,0){$T~~\sim~~$}
\multiframe(10,3)(10.5,0){2}(10,10){}{}
\multiframe(10,-7.5)(10.5,0){1}(10,10) {}
\put(35,0){$\longrightarrow$}\put(35,-10){\ft{$d^{\{1\}}$}}
\multiframe(60,3)(10.5,0){2}(10,10){}{}
\multiframe(60,-7.5)(10.5,0){1}(10,10){}
\multiframe(60,-18)(10.5,0){1}(10,10){\ft$\pa$}
\put(90,0){,}
\multiframe(110,3)(10.5,0){2}(10,10){}{}
\multiframe(110,-7.5)(10.5,0){1}(10,10) {}
\put(135,0){$\longrightarrow$}\put(135,-10){\ft{$d^{\{2\}}$}}
\multiframe(160,3)(10.5,0){2}(10,10){}{}
\multiframe(160,-7.5)(10.5,0){2}(10,10){}{\ft$\pa$}
\put(190,0){,}
\multiframe(210,3)(10.5,0){2}(10,10){}{}
\multiframe(210,-7.5)(10.5,0){1}(10,10) {}
\put(235,0){$\longrightarrow$}\put(235,-10){\ft{$d^{\{3\}}$}}
\multiframe(260,3)(10.5,0){3}(10,10){}{}{\ft$\pa$}
\multiframe(260,-7.5)(10.5,0){1}(10,10){}
\put(300,0){,}
\put(320,0){$etc.$}
\end{picture}

For fixed $i$ and $j$ the set of ghosts $A^{(i,j)}$ and all
their derivatives decompose into three types of independent
variables: \bqn [A^{(i,j)}]\quad \longleftrightarrow \quad \co
A^{(i,j+1)}\,,\,\co d^{\{1\}}A^{(i,j+1)}\,,\,\co
d^{\{2\}}A^{(i,j+1)}\,,\,\co d^{\{1\}}d^{\{2\}}A^{(i,j+1)} \nonumber \eqn
where $\co$ denotes any operator of the type $\prod_{m \geqslant
3}d^{\{m\}}\,$  or the identity.

Different cases arise depending on the position of the field
$A^{(i,j)}$ in Figure 1. We have to consider fields that sit in
the interior, on a border or at a corner of the diagram.

\begin{itemize}
\item[$\bullet$]\underline{Interior}

\noindent In this case, all the ghosts $A^{(i,j)}$ and their
derivatives form $u^{\ell}$ or $v^{\ell}$ variables. Indeed, we
have the relations
\begin{eqnarray}
\g  A^{(i,j)}&\propto&\big[d^{\{1\}}A^{(i+1,j)}+d^{\{2\}}A^{(i,j+1)}\big]\,,\nonumber\\
\g \big[d^{\{1\}}A^{(i+1,j)}-d^{\{2\}}A^{(i,j+1)}\big]&=&0\,,\nonumber\\
\g \big[d^{\{1\}}A^{(i+1,j)}+d^{\{2\}}A^{(i,j+1)}\big] &\propto& d^{\{1\}}d^{\{2\}}A^{(i,j+1)}\,,\nonumber\\
\g \big[d^{\{1\}}d^{\{2\}}A^{(i,j+1)}] &=&0
\,,\nonumber\end{eqnarray} and $\co$ commutes with $\gamma$. From
which we conclude that one can perform a change of variable from
the sets $[A^{(i,j)}]$ to the contractible pairs
\begin{eqnarray}
u^\ell &\leftrightarrow & \co A^{(i,j+1)}\,,\,\co
\big[d^{\{1\}}A^{(i,j+1)}+ d^{\{2\}}A^{(i,j+1)}]\nonumber\\
v^\ell &\leftrightarrow & \co [d^{\{1\}}A^{(i,j+1)}-
d^{\{2\}}A^{(i,j+1)}]\,,\,\co
d^{\{1\}}d^{\{2\}}A^{(i,j+1)}\nonumber
\end{eqnarray}so that the ghosts
$A^{(i,j)}$ in the interior and all their derivatives do not
appear in $H(\g)\,$.

\item[$\bullet$]\underline{Lower corner}

\noindent On the one hand, we have $\g A_{[q,0]}^{(p-q,q)}=0\,$.
As the operator $\g$ introduces a derivative, $
A_{[q,0]}^{(p-q,q)}\,$ cannot be $\g$-exact. As a result, $A_{[q,0]}^{(p-q,q)}$ is a
$w^i$-variable and thence belongs to $H(\g)\,$. On the other hand,
we find $\pa_{\n}A_{\m_1\ldots\m_q}^{(p-q,q)}=$
$\g\big[A^{(p-q-1,q)}_{\n\m_1\ldots\m_q}+$
$(-)^{p-q}\frac{q}{p+1}A^{(p-q,q-1)}_{\m_1\ldots\m_q\vert\,\n}\big]\,$,
which implies that all the derivatives of $A^{(p-q,q)}$ do not
appear in $H(\g)\,$. 

\item[$\bullet$]\underline{Border}

\noindent  If a ghost $A^{(i,j)}$ stands on a border of Figure 1,
it means that either (i) its reducibility relation involves
only one ghost (see e.g. Fig. 3), or (ii) there exists only one
field whose reducibility relation involves $A^{(i,j)}\,$ (see
e.g. Fig. 2):
\begin{itemize}
\item[(i)] Suppose $A^{(i,j)}$ stands on the left-hand (lower)
edge of Figure 1. We have the relations \begin{eqnarray}\g
A^{(i,j)}&\propto& d^{\{2\}}A^{(i,j+1)}\,,\nonumber\\
\g \big[d^{\{2\}}A^{(i,j+1)}\big]&=&0\,,\nonumber\\
\g \big[d^{\{1\}}A^{(i,j)}\big]&\propto&
d^{\{1\}}d^{\{2\}}A^{(i,j+1)}\,,\nonumber\\
\g \big[d^{\{1\}}d^{\{2\}}A^{(i,j+1)}\big]&=& 0\,,\nonumber
\end{eqnarray} so that the corresponding sets $[A^{(i,j)}]$ on the
left-hand edge do not contribute to $H(\g)$. We reach similar
conclusion if $A^{(i,j)}$ lies on the right-hand (higher) border
of Figure 1, substituting $d^{\{1\}}$ for $d^{\{2\}}$ when
necessary. \item[(ii)] Since, by assumption, $A^{(i,j)}$ does not
sit in a corner of Fig. 1 (but on the higher left-hand or lower
right-hand border), its reducibility transformation involves two
ghosts, and we proceed as if it were in the interior. The only
difference is that $\co d^{\{1\}}d^{\{2\}}A^{(i,j)}$ will be equal
to either $\g \co d^{\{1\}}A^{(i,j-1)}$ or $\g \co
d^{\{2\}}A^{(i-1,j)}\,$, depending whether the field above
$A^{(i,j)}$ is $A^{(i-1,j)}$ or $A^{(i,j-1)}\,$.
\end{itemize}

\item[$\bullet$]\underline{Left-hand corner}

\noindent In this case, the ghost $A^{(i,j)}$ is characterized by
a squared-shape Young diagram (it is the only one with this
property). Its reducibility transformation involves only one ghost
and there exists only one field whose reducibility transformation
involves $A^{(i,j)}\,$. Because of its symmetry properties,
$d^{\{2\}}A^{(i,j)}\sim d^{\{1\}}A^{(i,j)}\,$. Better,
$d^{\{2\}}$ is not well-defined on $A^{(i,j)}\,$, it is only
well-defined on $d^{\{1\}}A^{(i,j)}\,$. Therefore, the derivatives
$\pa_{\m_1\ldots\m_k}A^{(i,j)}$ decompose into $\co A^{(i,j)}\,$,
$\co d^{\{1\}}A^{(i,j)}\,$ and $\co
d^{\{1\}}d^{\{2\}}A^{(i,j)}\,$. The first set $\co A^{(i,j)}$ form
$u^{\ell}$-variables associated with $\co d^{\{2\}}A^{(i,j+1)}\,$.
The second set is grouped with $\co
d^{\{1\}}d^{\{2\}}A^{(i,j+1)}\,$, and the third one form
$v^{\ell}$-variables with $\co d^{\{2\}}A^{(i-1,j)}\,$.

\item[$\bullet$]\underline{Upper-corner}

\noindent In the case where $A^{(i,j)}$ is the gauge field, we proceed exactly as in the
``Interior'' case, except that the variables $\co d^{\{1\}}d^{\{2\}}A^{(i,j)}=0$ are not
grouped with any other variables any longer. They constitute true $w^i$-variables and are
thus present in $H(\g)\,$.
Recalling the definition of the curvature $K\,$, we have
$\co d^{\{1\}}d^{\{2\}}A^{(i,j)}\propto [K]\,$.

\item[$\bullet$]\underline{Right-hand corner}

\noindent In this case, the field $A^{(i,j)}$ is the $p$-form ghost $A^{(0,q)}_{[p]}\,$.
We have the $(u,v)$-pairs
$(\co d^{\{2\}}A^{(0,q)},\co d^{\{1\}}d^{\{2\}} A^{(1,q)})\,$,
$( \co d^{\{1\}}A^{(0,q-1)} , \co d^{\{1\}}d^{\{2\}} A^{(0,q)} )\,$.

\noindent The derivative $d^{\{1\}}A_{[p]}^{(0,q)}$ $\propto$ $D^0_{[p+1]}$ is a
$w^i$-variable since
it is invariant and no other variable $\pa_{\m_1\ldots \m_k}A^{(i,j)}$ possesses the
same symmetry.
\end{itemize}
$\Box$

In the sequel, the polynomials $\a ([K],[A^*])$ in the curvature, the antifields and all their derivatives will be called ``invariant polynomials''.
We will denote by ${\cal N}$ the algebra generated by all the ghosts
and the non-invariant derivatives of the field $\phi$.
The entire algebra of the fields and antifields is then generated by the invariant polynomials and the elements of ${\cal N}$.

\section{Invariant Poincar\'e lemma}
\label{InvariantPoincarelemma}

The space of {\it invariant} local forms is the space of (local)
forms that belong to $H(\gamma)$. The algebraic Poincar\'e lemma
 tells us that any closed form is exact.
However, if the form is furthermore invariant, it is not guaranteed
that the form is exact in the space of invariant forms. The
following lemma  tells us more about this important subtlety, in a
limited range of form degree.

\begin{lemma}[Invariant Poincar\'e lemma in form degree $k<p+1$]\label{invPoinclemma}
Let $\a^k$ be an invariant local $k$-form, $k<p+1\,$. 
$$\mbox{If}\quad d\a^k=0\,,\quad \mbox{then}
\quad\a^k=Q(K^{q+1}_{\m_1 \ldots \m_{p+1}})+d\b^{k-1}\,,$$ where
$Q$ is a polynomial in the $(q+1)$-forms $$K^{q+1}_{\m_1 \ldots
\m_{p+1}}\equiv K_{\m_1 \ldots \m_{p+1} \vert\, \n_1 \ldots
\n_{q+1}} dx^{\n_1} \ldots d x^{\n_{q+1}}\,,$$ while $\b^{k-1}$ is
an invariant local form. 

A closed invariant local form of form-degree $k<p+1$ and
of strictly positive antighost number is always exact in the space
of invariant local forms.
\end{lemma}

\noindent The proof is directly inspired from the one given in
\cite{Henneaux:1996ws} (Theorem 6).

\subsection{Beginning of the proof of the invariant Poincar\'e lemma}
\label{Beginningoftheproof}

The second statement of the lemma (\ie  the case $antigh(\a^k)\neq
0$) is part of a general theorem (see e.g. \cite{Dubois-Violette:1991is}) which holds without any restriction on the form-degree. It will not be reviewed here.

We will thus assume that $antigh(\a^k) =0$, and prove the first
part of Lemma \ref{invPoinclemma} by induction:
\begin{description}
    \item[Induction basis:] For $k=0$, the invariant Poincar\'e lemma is trivially satisfied: $d \a^0=0$ implies that $\a^0$ is a constant by the usual Poincar\'e lemma.
    \item[Induction hypothesis:] The lemma 
    holds in form degree $k^\prime$ such that $0\leqslant k^\prime <k$
    \item[Induction step:] We will prove in the sequel that under the induction
hypothesis, the lemma  holds in form degree
$k$. 

Because $d\a^k=0$ and $\gamma\a^k=0$, we can build a descent as
follows\begin{eqnarray}
d\a^k=0 \Rightarrow \a^k&=&da^{k-1,0} \label{form320asub}\\
0&=&\gamma a^{k-1,0} + da^{k-2,1}\label{form321asub}\\
& \vdots& \nonumber\\
0&=&\gamma a^{k-j,j-1} + da^{k-j-1,j}\label{form323asub} \\
0&=&\gamma a^{k-j-1,j}\,,\label{form324asub}
\end{eqnarray}
where $a^{r,i}$ is a $r$-form of pureghost number $i\,$.
The pureghost number of  $a^{r,i}$ must obey
$0\leqslant i\leqslant k-1\,$.
Of course, since we assume $k<p+1\,$, we have $i<p\,$.
The descent stops at (\ref{form324asub}) either because $k-j-1=0$ or
because $a^{k-j-1,j}$ is invariant.
The case $j=0$ is trivial since it gives immediately $\a^k=d\b^{k-1}\,$,
where $\b^{k-1}\equiv a^{k-1,0}$ is invariant. Accordingly, we assume from now
on that $j>0\,$.

Since we are dealing with a descent, it is helpful to introduce one of its
building blocks, which is the purpose of the next subsection. We will
complete the induction step in Subsection \ref{endprfindstp}.
\end{description}

\subsection{A descent of $\g$ modulo $d$}\label{descentogmodulod}
\label{descentgammamodd}

Let us define the following differential forms built up from the
ghosts
$$D^l_{\m_1 \ldots \m_{p+1}} \equiv (-)^{l(q+1)+q}\pa^{}_{[\m_1}
A^{(0,q-l)}_{\m_2 \ldots \m_{p+1} ] \vert\, \n_1 \ldots \n_{l}}
dx^{\n_1} \ldots d x^{\n_l}\,,$$ for $0 \leqslant l \leqslant q\,$.
It is easy to show that these fields verify the following descent:
\bqn
\g (D^0_{\m_1 \ldots \m_{p+1}})&=&0  \,, \label{desc1} \\
\g (D^{l+1}_{\m_1 \ldots \m_{p+1}})+d D^l_{\m_1 \ldots \m_{p+1}}&= &0 \;,\quad\quad 0 \leqslant l \leqslant q-1\,,\nonumber \\
d D^q_{\m_1 \ldots \m_{p+1}}&=&K^{q+1}_{\m_1 \ldots \m_{p+1}}\,.
\label{desc2}
\eqn
It is convenient to introduce the inhomogeneous form
$$D_{\m_1\ldots \m_{p+1}}=\sum_{l=0}^{q}D^l_{\m_1 \ldots \m_{p+1}} $$
because it satisfies a so-called ``Russian formula" \be (\g +
d)D_{\m_1 \ldots \m_{p+1}}=K^{q+1}_{\m_1 \ldots \m_{p+1}}\,,
\label{russianf} \ee which is a compact way of writing the descent
(\ref{desc1})--(\ref{desc2}). \vspace*{.2cm}

Let $\o_{(n,m)}$ be a homogeneous polynomial of degree $m$ in
$D$ and of degree $n$ in $K$. Its decomposition is
$$\o_{(n,m)}(K,D)=\o^{n(q+1)+mq,0}+...+\o^{n(q+1)+j,mq-j}+...+\o^{n(q+1),mq}$$ where $\o^{n(q+1)+j,mq-j}$ has form degree $n(q+1)+j$
and pureghost number $mq-j$.  Due to (\ref{russianf}), the
polynomial satisfies \bqn (\g + d)\o_{(n,m)}=K^{q+1}_{\m_1 \ldots
\m_{p+1}} \frac{\partial^L \o_{(n,m)}}{\partial D_{\m_1 \ldots
\m_{p+1}}}\,,\label{rhsdd}\eqn the form degree decomposition of
which leads to the descent 
\bqn
\g (\o^{n(q+1),mq})&=& 0\,,\nonumber \\
\g (\o^{n(q+1)+j+1,mq-j-1})+d \o^{n(q+1)+j,mq-j} &= & 0\, ,\quad0
\leqslant  j \leqslant q-1\nonumber
\\
\g (\o^{n(q+1)+q+1,(m-1)q-1}) + \,d \o^{n(q+1)+q,(m-1)q} &= &
K_{\m_1 \ldots \m_{p+1}}^{q+1}\Big[\frac{\partial^L \o}{\partial
D_{\m_1 \ldots \m_{p+1}}}\Big]^{n(q+1),(m-1)q} \label{rhsd} \eqn
where $[\frac{\partial \o}{\partial D}]^{n(q+1),(m-1)q}$ denotes
the component of form degree $n(q+1)$ and pureghost equal to
$(m-1)q$ of the derivative $\frac{\partial \o}{\partial D}$. 
This
component is the homogeneous polynomial of degree $m-1$ in the
variable $D^0$,  $$\Big[\frac{\partial \o}{\partial D_{\m_1
\ldots \m_{p+1}}}\Big]^{n(q+1),(m-1)q}=\frac{\partial \o}{\partial
D_{\m_1 \ldots \m_{p+1}}}\vert_{D=D^0}\,.$$ The
right-hand-side of (\ref{rhsd}) vanishes if and only if the
right-hand-side of (\ref{rhsdd}) does.

Two cases arise depending on whether the r.h.s. of (\ref{rhsdd}) vanishes or not.
\begin{itemize}
    \item The r.h.s. of (\ref{rhsdd}) vanishes: then the descent is said not to be obstructed in any strictly positive pureghost number and goes all the way down
    to the bottom equations\begin{eqnarray}
\g (\o^{n(q+1)+mq,0})+d \o^{n(q+1)+mq+1,1} &= & 0\, ,\quad0
\leqslant j \leqslant q-1\nonumber
\\ d(\o^{n(q+1)+mq,0})&=& 0\,.\nonumber
\end{eqnarray}
    \item The r.h.s. of (\ref{rhsdd}) is not zero : then the descent is  obstructed
    after $q$ steps. 
It is not possible to find an $\tilde{\o}^{n(q+1)+q+1,(m-1)q-1} $ such that $$\g (\tilde{\o}^{n(q+1)+q+1,(m-1)q-1} )+ \,d \o^{n(q+1)+q,(m-1)q} = 0\,,$$
because the r.h.s. of (\ref{rhsd}) is  an element of
    $H(\gamma)$. This element is
    called the {\it obstruction} to the descent. One also says   that this obstruction
    cannot be lifted more than $q$ times, and $\o^{n(q+1),mq}$ is
    the top of the ladder (in this case it must be an element of $H(\gamma)$).
\end{itemize}
This covers the general type of ladder (descent as well as lift)
that do not contain the $p\,$-th generation ghost $A^{(p-q,q)}$.

\subsection{End of the proof of the invariant Poincar\'e lemma}
\label{endprfindstp}

As $j<p$, Theorem \ref{Hgamma} implies that the equation
(\ref{form324asub}) has non-trivial solutions only when $j=mq$ for
some integer $m$
\begin{equation}
a^{k-mq-1,mq}=\sum_I\a_I^{k-mq-1}\,\o_I^{0,mq}\,, \label{blbl}
\end{equation}
up to some $\gamma$-exact term. The $\a_I^{k-mq-1}$'s are invariant
forms, and $\{\o_I^{0,mq}\}$ is a basis of polynomials of degree
$m$ in the variable $D^0$. The ghost $A^{(p-q,q)}$ are absent
since the pureghost number is $j=mq< p$.

The equation (\ref{form323asub}) implies $d\a_I^{k-mq-1}=0$. Together with the induction hypothesis, this implies
\begin{equation}
\a_I^{k-mq-1}=P_I(K^{q+1}_{\m_1 \ldots
\m_{p+1}})+d\b^{k-j-2}\quad\,,\label{blblbl}
\end{equation}where the polynomials $P_I$ of order $n$ are present
iff $k-mq-1=n(q+1)$. Inserting (\ref{blblbl}) into (\ref{blbl}) we
find that, up to trivial redefinitions,
$a^{k-j-1,j}$ is a polynomial in $K^{q+1}_{\m_1 \ldots \m_{p+1}}$
and $D^0_{\,\m_1 \ldots \m_{p+1}}$.

From the analysis performed in Subsection \ref{descentogmodulod},
we know that such an $a^{k-j-1,j}$ can be lifted at most $q$
times. Therefore, $a^{k-j-1,j}$  belongs to a descent of
type (\ref{form320asub})--(\ref{form324asub}) only if $j=q\,$.
 Without loss of generality we can thus take
$a^{k-q-1}_q=P(K^{q+1}_{\m_1 \ldots \m_{p+1}},D^0)$ where $P$ is
a homogeneous polynomial with a linear dependence in $D^0$ (since
$m=1$). In such a case, it can be lifted up to (\ref{form320asub}). Furthermore, because $a^{k-1,0}$ is defined
up to an invariant form $\b^{k-1,0}$ by the equation
(\ref{form321asub}), the term $da^{k-1,0}$ of (\ref{form320asub})
must be equal to the sum
$$da^{k-1,0}=\underbrace{P(K^{q+1},K^{q+1})}_{\equiv Q(K^{q+1}_{\m_1 \ldots \m_{p+1}})}+d\b^{k-1,0}$$
of a homogeneous polynomial $Q$ in $K^{q+1}$ (the lift of the
bottom) and a form $d$-exact in the invariants. $\qedsymbol$

\section{Cohomology of $\d$ modulo $d\,$: $H^D_k(\d \vert\, d)$}
\label{Characteristiccohomology}

In this section, we compute the cohomology of $\d$ modulo $d$ in
top form-degree and antighost number $k$, for $k\geqslant q\,$.
We will also restrict ourselves to $k>1\,$. The group $H^D_1(\d \vert\, d)$ describes the infinitely many conserved currents and will not be studied here.
\vspace*{.2cm}

Let us first recall a general  theorem (Theorem 9.1 in \cite{Barnich:db}).

\begin{theorem}\label{usefll}
For a linear gauge theory of reducibility order $p-1$, \bqn
H_k^D(\d \vert\, d)=0\; for\; k>p+1\,. \nonumber\eqn
\end{theorem}

The computation of the cohomology groups $H_k^D(\d \vert\, d)$
for $q \leqslant k\leqslant p+1$ follows closely the procedure  used for
$p$-forms in \cite{Henneaux:1996ws}. It relies on the following
theorems:

\begin{theorem}\label{bilin}
Any solution of $\d a^D+d b^{D-1} =0$ that is at least bilinear in the antifields is necessarily
trivial.
\end{theorem}
The proof of Theorem \ref{bilin} is similar to the proof of Theorem 11.2 in
\cite{Barnich:db} and will not be repeated here.

\begin{theorem}\label{pplusun}
A complete set of representatives of $H^D_{p+1}(\d \vert\, d)$ is
given by the antifields $C^{* \, D}_{p+1\,\m_1 \ldots \m_q}$, i.e.
\bqn
 \d a^D_{p+1}+d a^{D-1}_{p}=0
 \;\Rightarrow \; a^D_{p+1}=\l^{\m_{[q]}}C^{* \, D}_{p+1\,\m_{[q]}} +
\d b_{p+2}^D+d b_{p+1}^{D-1}\,,
 \nonumber
 \eqn
where the $\l^{[\m_1 \dots \m_q]}$ are constants.
\end{theorem}
\proof{ \underline{Candidates}: any polynomial of antighost
number $p+1$ can be written \bqn a_{p+1}^D=\Lambda^{[\m_1 \dots
\m_q]}C^{* \, D}_{p+1\,[\m_1 \dots \m_q]} +\m_{p+1}^D+\d
b_{p+2}^D+d b_{p+1}^{D-1}\,, \nonumber \eqn where $\Lambda$ does
not involve the antifields and where $\m_{p+1}^D$ is at least
quadratic in the antifields. The cocycle condition $ \d
a^D_{p+1}+d a^{D-1}_{p}=0$ then implies \bqn -\Lambda^{[\m_1 \dots
\m_q]}d C^{* \, D-1}_{p\,[\m_1 \dots \m_q]}+ \d(\m_{p+1}^D +d
b_{p+1}^{D-1})=0\,. \nonumber \eqn By taking the Euler-Lagrange
derivative of this equation with respect to $C^{*}_{p\,[\m_1 \dots
\m_q]\vert\, \n}$, one gets the weak equation
$\pa^{\n}\Lambda^{[\m_1 \dots \m_q]}\approx 0\,$. Considering
$\n$ as a form index, one sees that $\Lambda$ belongs to
$H_0^0(d\vert\, \d)$. The isomorphism $H_0^0(d\vert\, \d)/
\mathbb{R} \cong H^D_{D}(\d\vert\, d)$ (see \cite{Barnich:db})
combined with the knowledge of $H^D_{D}(\d\vert\, d)\cong 0$ (by
Theorem \ref{usefll}) implies $\Lambda^{[\m_1 \dots
\m_q]}=\l^{[\m_1 \dots \m_q]}+\d \n_1^{[\m_1 \dots \m_q]}$ where
$\l^{[\m_1 \dots \m_q]}$ is a constant. The term $\d \n_1^{[\m_1
\dots \m_q]}C^{* \, D}_{p+1\,[\m_1 \dots \m_q]} $ can be rewritten as a term at least bilinear in the antifields up to a $\d$-exact
term.  Inserting $a_{p+1}^D=\l^{[\m_1 \dots \m_q]}C^{* \,
D}_{p+1\,\m_1 \dots \m_q}+\m_{p+1}^D+\d b_{p+2}^D+d b_{p+1}^{D-1}$
into the cocycle condition, we see that $\m_{p+1}^D$ has to be a
solution of $\d \m_{p+1}^D+d b^{D-1}=0$ and is therefore
trivial by Theorem \ref{bilin}. \vspace*{.2cm}

\underline{Non-triviality}: It remains to show that the
cocycles $a^D_{p+1}=\l C^{*\,D}_{p+1}$ are non-trivial. Indeed one
can prove that  $\l C^{*\,D}_{p+1}=\d u_{p+2}^D + d v_{p+1}^{D-1}$
implies that
 $\l C^{*\,D}_{p+1}$ vanishes. It is straightforward when $u_{p+2}^D$ and $v_{p+1}^{D-1}$
do not depend explicitly on $x$: $\d$ and $d$ bring in a
derivative while $\l C^{*\,D}_{p+1}$ does not contain any. If $u$
and $v$ depend explicitly on $x$, one must expand them and the
equation $\l C^{*\,D}_{p+1}=\d u_{p+2}^D + d v_{p+1}^{D-1}$
according to the number of derivatives of the fields and
antifields to reach the conclusion. Explicitly, $u_{p+2}^D
=u_{p+2,\,0}^D + \ldots +u_{p+2,\,l}^D$ and $
v_{p+1}^{D-1}=v_{p+1,\,0}^{D-1}+ \ldots +v_{p+1,\,n}^{D-1}$. If
$n>l$, the equation in degree $n+1$ reads $0= d'
v_{p+1,\,n}^{D-1}$ where $d'$ does not differentiate with respect
to the explicit dependence in $x$. This in turn implies that
$v_{p+1,\,n}^{D-1}= d' \tilde{v}_{p+1,\,n-1}^{D-1}$ and can be
removed by redefining $v_{p+1}^{D-1}$: $v_{p+1}^{D-1} \rightarrow
v_{p+1}^{D-1}-d \tilde{v}_{p+1,\,n-1}^{D-1}$. If $l>n$,  the
equation in degree $l+1$ is $0=\d u_{p+2,\,l}^D$ and implies,
together with the acyclicity of $\d$, that one can remove
$u_{p+2,\,l}^D$ by a trivial redefinition of $u_{p+2}^D\,$. If
$l=n>0$, the equation in degree $l+1$ reads $0= \d u_{p+2,\,l}^D+
d' v_{p+1,\,l}^{D-1}\,$. Since there is no cohomology in antighost
number $p+2$, this implies that $u_{p+2,\,l}^D=\d
\bar{u}_{p+3,\,l-1}^D + d'\tilde{u}_{p+2,\,l-1}^{D-1}$ and can be
removed by trivial redefinitions: $u_{p+2}^D \rightarrow u_{p+2}^D
-\d \bar{u}_{p+3,\,l-1}^D $ and $v_{p+1}^{D-1}\rightarrow
v_{p+1}^{D-1}- d \tilde{u}_{p+2,\,l-1}^{D-1}\,$. Repeating the
steps above, one can remove all $u_{p+2,\,l}^D$ and
$v_{p+1,\,n}^{D-1}$ for $l,\,n>0\,$. One is left with $\l
C^{*\,D}_{p+1}=\d u_{p+2,\,0}^D + d' v_{p+1,\,0}^{D-1}\,$. The
derivative argument used in the case without explicit
$x$-dependence now leads to the desired conclusion.}

\begin{theorem}\label{BKzero}
The cohomology groups $H_k^D(\d \vert\, d)$ ($k>1$) vanish
unless $k=D-r(D-p-1)$ for some strictly positive integer $r\,$.
Furthermore, for those values of $k\,$,  $H_k^D(\d \vert\, d)$ 
has at most one non-trivial class.
\end{theorem} 
\proof{We already know that $H_k^D(\d \vert\,
d)$ vanishes for $k>p+1$ and that $H_{p+1}^D(\d \vert\, d)$ has one non-trivial class. Let us assume that the theorem has been proved for
all $k$'s strictly greater than $K$ (with $K<p+1$) and  extend it to $K$.
Without loss of generality we can assume that the cocycles of
$H_K^D(\d \vert\, d)$ take the form (up to trivial terms)
$a_{K}=\l^{\m_1 \ldots \m_{p+1-K}\vert\, \n_1 \ldots \n_q}C^*_{K\;
\n_1 \ldots \n_q \vert\, \m_1 \ldots \m_{p+1-K}}+ \m $, where $\l$
does not involve the antifields and $\m$ is at least bilinear in
the antifields. Taking the Euler-Lagrange derivative of the
cocycle condition with respect to $C^*_{K-1}$ implies that
$\l^{p+1-K}_{\n_1 \ldots \n_q}\equiv \l_{\m_1 \ldots
\m_{p+1-K}\vert\, \n_1 \ldots \n_q}dx^{\m_1} \ldots
dx^{\m_{p+1-K}}$ defines an element of $H_0^{p+1-K}(d \vert\,
\d)$. If $\l$ is $d$-trivial modulo $\d$, then it is
straightforward to check that $\l  C^{*\,D-p-1+K}_{K}$ is trivial
or bilinear in the antifields. Using the isomorphism
$H_0^{p+1-K}(d \vert\, \d) \cong H_{D-p-1+K}^{D}(\d \vert\, d)$,
we see that $\l$ must be trivial unless $ D-p-1+K=D-r(D-p-1)\,$, in which case $H_{D-p-1+K}^{D}(\d \vert\, d)$ has one non-trivial class. Since $K=D-(r+1)(D-p-1)$ is also of the required form, the
theorem extends to $K$. }

\begin{theorem} \label{BK}
Let $r$ be a strictly positive integer. A complete set of
representatives of $H_k^D(\d \vert\, d)$ ($k=D-r(D-p-1)\geqslant q
$) is given by the terms of form-degree $D$ in the expansion of all possible homogeneous
polynomials $P(\tilde{H})$ of degree $r$ in $\tilde{H}$ (or equivalently $P(\tilde{\cal H})$ of degree $r$ in $\tilde{\cal H}$). 
\end{theorem}
\noindent The proof of this theorem is given in Appendix
\ref{ProofoftheoremBK}. \vspace{.2cm}

These theorems give us a complete description of all the
cohomology group $H_k^D(\d\vert\, d)$ for $k\geqslant q $ (with $k>1\,$).

\section{Invariant cohomology of $\d$ modulo $d$, $H_k^{inv}(\d \vert\, d)$}
\label{Invariantcharacteristiccohomology}

In this section, we compute the set of invariant  solutions $a^D_k$ ($k \geqslant q$) of the
equation $\d a^D_k+d b^{D-1}_{k-1}=0$, up to trivial terms
$a^D_k=\d b^D_{k+1}+d c^{D-1}_k$, where $b^D_{k+1}$ and
$c^{D-1}_k$ are invariant. This space of solutions is the
invariant  cohomology of $\d$ modulo $d$, $H_k^{inv}(\d \vert\, d)$. 
We first  compute representatives of all the  cohomology classes of $H_k^{inv}(\d \vert\, d)$, then we find out the cocycles without explicit $x$-dependance.

\begin{theorem} \label{cohoinva}
For $k\geqslant q$, a complete set of invariant solutions  of the equation $\d a^D_k+db^{D-1}_{k-1}=0$ is given by the polynomials in the curvature $K^{q+1}$ and in $\tilde{\cal H}$ (modulo trivial solutions): 
$$\d a^D_k+d b^{D-1}_{k-1}=0 \Rightarrow a^D_k= P(K^{q+1},\tilde{\cal H})\vert\,_k^D +\d \m_{k+1}^D+d \n_k^{D-1} \,,$$
where $\m_{k+1}^D$ and $\n_k^{D-1}$ are invariant forms.
\end{theorem}

\proof{
From the previous section, we know that for $k\geqslant q$ the
general solution of the equation $\d a^D_k+d
b^{D-1}_{k-1}=0$ is $a^D_k=Q(\tilde{\cal H})\vert\,^D_k+\d m^D_{k+1}+d
n^{D-1}_k$ where $Q(\tilde{\cal H})$ is a homogeneous polynomials of
degree $r$ in $\tilde{\cal H}$ (it exists only when $k=D-r(D-p-1)$).
Note  that $m^D_{k+1}$ and $n^{D-1}_k$ are not necessarily
invariant. However, one can prove the following theorem (the lengthy proof of which is given in the
Appendix \ref{Proofoftheoremcohoinv}):

\begin{theorem}\label{cohoinv}
Let $\a^D_k$ be an invariant polynomial ($k\geqslant q$). If $\a_k^D = \d
m_{k+1}^D + d n_k^{D-1} $, then
$$\a_k^D = R^{(s,r)}(K^{q+1},\tilde{\cal H})\vert\,_k^D+\d \m_{k+1}^D + d \n_k^{D-1}\,,$$
where  $R^{(s,r)}(K^{q+1},\tilde{\cal H})$ is a polynomial of degree $s$ in $K^{q+1}$ and $r$ in $\tilde{\cal H}$, such that the strictly positive integers $s,r$ satisfy
 $D=r(D-p-1)+k+s(q+1)$ and $\m_{k+1}^D$ and $\n_k^{D-1}$ are invariant forms.
\end{theorem}

\noindent As $a^D_k$ and $Q(\tilde{\cal H})\vert\,^D_k$ are invariant, this theorem implies that
$$a^D_k=P^{(s,r)}(K^{q+1},\tilde{\cal H})\vert\,_k^D+\d \m_{k+1}^D + d \n_k^{D-1}\,,$$
 where $P^{(s,r)}(K^{q+1},\tilde{\cal H})$ is a polynomial of non-negative degree $s$ in $K^{q+1}$ and of strictly positive degree $r$ in $\tilde{\cal H}$. Note that the polynomials of non-vanishing degree in $K^{q+1}$ are trivial in $H_k^{D}(\d \vert\, d)$ but not necessarily in $H_k^{D\, inv}(\d \vert\, d)$. }

Part of the  solutions found in Theorem \ref{cohoinva} depend explicitely on the coordinate $x$, because $\tilde{\cal H}\vert \,_0$ does.  Therefore the question arises
whether there exist other representatives of the same
non-trivial equivalence class $[P^{(s,r)}(K^{q+1},\tilde{\cal H})\vert\,^D_k]\in
H^{D\, inv}_k(\d\mid d)$ that \textit{do not} depend explicitly on $x$. The answer is
negative when $r>1$.  In other words, we can prove the general theorem:

\begin{theorem}\label{quad}
When $r>1$, there is no non-trivial invariant cocycle in the equivalence class $[P^{(s,r)}(K^{q+1},\tilde{\cal H})\vert\,^D_k]$ $\in H^{D\, inv}_k(\d\mid d)$ without explicit $x$-dependance. 
\end{theorem}

To do so, we first prove the following lemma:
\begin{lemma}Let $P(K^{q+1},\tilde{\cal H})$ be a homogeneous polynomial of order $s$ in the curvature $K^{q+1}$ and $r$ in $\tilde{\cal H}$.
If $r\geqslant 2$, then the component $P(K^{q+1},\tilde{\cal H})|^D_k$ always contain terms of
order $r-1$($\neq 0$) in $\tilde{\cal H}\vert \,_0 $. \end{lemma} \proof{Indeed,
$P(K^{q+1},\tilde{\cal H})$ can be freely expanded in terms of $\tilde{\cal H}\vert \,_0 $ and
the undifferentiated antighost forms. The Grassmann parity is the
same for all terms in the expansion of $\tilde{\cal H}$, therefore the
expansion is the binomial expansion up to the overall coefficient
of the homogeneous polynomial and up to relative signs obtained
when reordering all terms. Hence, the component
$P(K^{q+1},\tilde{\cal H})|^D_k$ always contains a term that is a product of
$(r-1)$ $\tilde{\cal H}\vert \,_0^{ D-p-1}$'s, a single antighost $C^{*\,D-p-1+k}_k$ and $s$ curvatures,
which possesses the correct degrees as can be checked
straightforwardly.}

\noindent {\bf Proof of Theorem \ref{quad}:} \hspace{.5cm} Let us
assume that there exists a non-vanishing invariant $x$-independent
representative $\a^{D\,,\,inv}_{k}$ of the equivalence class
$[P^{(s,r)}(K^{q+1},\tilde{\cal H})|^D_k]\in H^{D \, inv}_k(\d\mid d)$, \ie \bqn
P^{(s,r)}(K^{q+1},\tilde{\cal H})|^D_k+\d\rho^{D}_{k+1}+d\sigma^{D-1}_{k}=\a^{D\,,\,inv}_{k}\,,\label{previousequ}\eqn
where $\rho^{D}_{k+1}$ and $\sigma^{D-1}_{k}$ are invariant and allowed to
depend explicitly on $x$.

We define the descent map $f:\a_m^n\rightarrow \a_{m-1}^{n-1}$
such that $\d \a_m^n+d \a_{m-1}^{n-1}=0$, for $n\leqslant D$. This
map is well-defined on equivalence classes of $H^{inv}(\d\vert d)$ when $m>1$ and
preserves the $x$-independence of a
representative. Hence, going down $k-1$ steps, it
is clear that the equation (\ref{previousequ}) implies:
$$
P^{(s,r)}(K^{q+1},\tilde{\cal H})|^{D-k+1}_1+\d\rho^{D-k+1}_2+d\sigma^{D-k}_1=\a^{D-k+1\,,\,inv}_1\,,$$
with $\a^{D-k+1\,,\,inv}_1\neq 0$.

We can decompose this equation in the polynomial degree in the
fields, antifields, and all their derivatives. Since $\d$ and $d$
are linear operators, they preserve this degree; therefore \be
P^{(s,r)}(K^{q+1},\tilde{\cal H})|^{D-k+1}_{1,\,r+s}+\d\rho^{D-k+1}_{2,\,r+s}+d\sigma^{D-k}_{1,\,r+s}
=\a^{D-k+1\,,\,inv}_{1,\,r+s}\,,\label{decompose}\ee where $r+s$
denotes the polynomial degree. The homogeneous polynomial
$\a^{D-k+1\,,\,inv}_{1,\,r+s}$ of polynomial degree $r+s$ is linear in
the antifields of antighost number equal to one, and depends on
the fields only through the curvature.

Finally, we introduce the number operator $N$ defined by \bqn
N\,&=&\,r\,\,\partial_{\r_1}\ldots\partial_{\r_r}\phi_{\m_1\ldots\m_p\,|\,\n_1\ldots\n_q}
\,\,\frac{\partial}{\partial
(\partial_{\r_1}\ldots\partial_{\r_r}\phi_{\m_1\ldots\m_p\,|\,\n_1\ldots\n_q})}\nn\\
&&+\,(r+1)\,\partial_{\r_1}\ldots\partial_{\r_r}\Phi_A^*\,\,\frac{\partial}{\partial
(\partial_{\r_1}\ldots\partial_{\r_r}\Phi_A^*)}
- x^{\m} \,\,\frac{\partial}{\partial x^{\m}}\nonumber 
\eqn
 where
$\{\Phi_A^*\}$ denotes the set of all antifields. It follows
immediately that $\d$ and $d$ are homogeneous of degree one and
the degree of $\tilde{\cal H}$ is also equal to one,
$$N(\d)=N(d)=1=N(\tilde{\cal H})\,.$$
Therefore, the decomposition in $N$-degree of the equation
(\ref{decompose}) reads in $N$-degree  equal to $n=r+2s$, \be
P^{(s,r)}(K^{q+1},\tilde{\cal H})|^{D-k+1}_{1,\,r+s}+\d\rho^{D-k+1}_{2,\,r+s,\,r+2s-1}+d\sigma^{D-k}_{1,\,r+s,\,r+2s-1}
=\a^{D-k+1\,,\,inv}_{1,\,r+s,\,r+2s}\label{endproof}\ee and, in $N$--degree
 equal to $n>r+2s$,
$$\d\rho^{D-k+1}_{2,\,r+s,\,n-1}+d\sigma^{D-k}_{1,\,r+s,\,n-1}
=\a^{D-k+1\,,\,inv}_{1,\,r+s,\,n}\,.$$ The component
$\a^{D-k+1\,,\,inv}_{1,\,r+s,\,r+2s}$ of $N$-degree  equal to $r+2s$ is $x$-independent, depends linearly on the (possibly differentiated) antighost of antifield number 1, and is of order $r+s-1$ in the (possibly differentiated) curvatures. Direct counting shows that there is no polynomial of $N$-degree equal to $r+2s$ satisfying these requirements when $r\geqslant 2$.
 Thus for $r\geqslant 2$ the component
$\a^{D-k+1\,,\,inv}_{1,\,r+s,\,r+s}$ vanishes, and then the equation
(\ref{endproof}) implies that $P^{(s,r)}(K^{q+1},\tilde{\cal H})|^{D-k+1}_{1,\,r+s}$ is trivial (and even vanishes when $s=0$, by Theorem \ref{BK}).

In conclusion, if $P(K^{q+1},\tilde{\cal H})$ is a polynomial that is quadratic or more in $\tilde{\cal H}$,
then there exists no non-trivial invariant representative without explicit $x$-dependence in the cohomology
class $[P(K^{q+1},\tilde{\cal H})]$ of $H^{inv}(\d\vert d)$. 
$\qedsymbol$
\vspace{.2cm}

This leads us to the following theorem:
\begin{theorem}
\label{thm8.1}
The invariant solutions $a^D_k$  ($k\geqslant q$) of the equation $\d a^D_k+d
b^{D-1}_{k-1}=0$ without explicit $x$-dependence are all trivial
in $H_k^{inv}(\d \vert\, d)$ unless $k=p+1-s(q+1)$ for
some non-negative  integer $s$. For those values of $k$, the
 non-trivial representatives are given by  polynomials that
are linear in $C^{*\; D-p-1+k}_k$ and of order $s$ in $K^{q+1}$. 
\end{theorem}

\proof{
By Theorem \ref{cohoinva}, invariant solutions of the equation $\d a^D_k+d
b^{D-1}_{k-1}=0$ are polynomials in $K^{q+1}$ and $\tilde{\cal H}$ modulo trivial terms. When the polynomial is quadratic or more in $\tilde{\cal H}$,  then Theorem \ref{quad} states that there is no representative without explicit $x$-dependance in its cohomology class, which implies that it should be rejected. The remaining solutions are the polynomials linear in  $\tilde{\cal H}\vert \,_k=C^{*\; D-p-1+k}_k$ and of arbitrary order in $K^{q+1}$. They are invariant and $x$-independent, they thus belong to the set of  looked-for solutions.
}

\section{Self-interactions}
\label{self-interactions}

As explained in Section \ref{BRST}, the non-trivial first order deformations
of the free theory are given by the elements of
$H^{D,\,0}(s\vert\, d)$, the cohomological group of the BRST differential $s$ in the space of local functionals 
in top form degree and in ghost number zero. The purpose of this
section is  to compute this group. As the computation is very
similar to the computation of similar groups in the case of
$p$-forms \cite{Henneaux:1997ha}, gravity\cite{Boulanger:2000rq},
dual gravity \cite{Bekaert:2002uh} and $[p,p]$-fields
\cite{Boulanger:2004rx}, we will not reproduce it here entirely
and refer to the works just cited (e.g. \cite{Boulanger:2000rq})
for technical details.  We just present the main steps of the
procedure and the calculations that are specific to the case of
$[p,q]$-fields. \vspace*{.2cm}

The proof is given for a single $[p,q]$-field $\phi$ but
extends trivially to a set $\{\phi^a\}$ containing a finite
number $n$ of them (with fixed $p$ and $q$) by writing some
internal index $a=1,\ldots,n$ everywhere. 
\vspace*{.2cm}

The group $H(s\vert\, d)$ is the group of solutions  $a$ of the
equation $sa+db=0$, modulo trivial solutions of the form $a= sm +
dn$. The basic idea to compute such a group is to use homological
perturbation techniques by expanding the quantities and the
equations according to the antighost number.

Let $a^{D,\,0}$ be a solution of $sa^{D,\,0}+db^{D-1,1}=0$ with
ghost number zero and top form degree. For convenience, we will
frequently omit to write the upper indices. One can expand
$a$($=a^{D,\,0})$ as $a=a_0+a_1 + \ldots + a_k$ where $a_i$
has antighost number $i$. The expansion can be assumed to stop at
some finite value of the antighost number under the sole
hypothesis that the first-order deformation of the Lagrangian has
a finite derivative order \cite{Barnich:mt}. Let us recall
\cite{Barnich:vg} that (i) the antifield-independent piece $a_0$
is the deformation of the Lagrangian; (ii) the terms linear in the
ghosts contain the information about the deformation of the
reducibility conditions; (iii) the other terms give the
information about the deformation of the gauge algebra.

Under the assumption of locality, the expansion of $b$ also stops at some finite antighost number.
Without loss of generality, one can assume that $b_j=0$ for
$j\geqslant k$. Decomposing the BRST differential as $s=\g + \d$,
the equation $sa+db=0$ is equivalent to \bqn
\d a_1 + \g a_0 + d b_0 &=&0 \,,\nonumber \\
\d a_2 + \g a_1 + d b_1 &=&0 \,,\nonumber \\
&\vdots &\nonumber \\
\d a_k + \g a_{k-1} + d b_{k-1}&=&0 \,,\nonumber \\
\g a_k&=&0 \,.\label{descente} \eqn 

The next step consists in the analysis of the term $a_k$ with highest antighost
number and the determination of whether it can be removed by trivial
redefinitions or not. We will see in the sequel under which assumptions
this can be done. \vspace*{.2cm}

\subsection{Computation of $a_k$ for $k>1$}

The last equation of the descent (\ref{descente}) is  $\g a_k =0$.
It  implies that $a_k=\a_J\,\o^J$ where $\a_J$ is an invariant
polynomial and $\o^J$ is a polynomial in the  ghosts of $H(\g)$:
$A^{(p-q,q)}_{\m_{[q]}}$ and $D^0_{\m_{[p+1]}}$. Inserting
this expression for $a_k$ into the second to last equation leads
to the result that $\a_J$ should be an element of
$H^{D,\,inv}_k(\d\vert\, d)$. Furthermore, if $\a_J$ is trivial in
this group, then $a_k$ can be removed by trivial redefinitions.
The vanishing of $H^{D,\,inv}_k(\d\vert\, d)$ is thus a sufficient
condition to remove the component $a_k$ from $a$. It is however
not a necessary condition, as we will see in the sequel.

We showed that non-trivial interactions can  arise only if some
$H^{D,\,inv}_k(\d\vert\, d)$ do not vanish. The requirement that
the Lagrangian should not depend explicitly on $x$ implies
that we can restrict ourselves to $x$-independent elements of
this group. Indeed, it can be shown \cite{book} that, when $a_0$ does not
depend explicitly on $x$,  the whole cocycle $a=a_0+a_1 +
\ldots + a_k$ satisfying $sa+db=0$ is $x$-independent  (modulo
trivial redefinitions). By Theorem \ref{thm8.1},
$H^{D,\,inv}_k(\d\vert\, d)$ contains non-trivial
$x$-independent elements only if $k=p+1-s(q+1)$ for some
non-negative integer $s$. The form of the non-trivial elements is then
$ \a_k^D=C_k^{*\,D-p-1+k} (K^{q+1})^s \,.$ In order to be
(possibly) non-trivial, $a_k$ must thus be a polynomial linear in
$C_k^{*\,D-p-1+k}$, of order $s$ in the curvature $K^{q+1}$ and of
appropriate orders in the ghosts $A^{(p-q,q)}_{\m_{[q]}}$ and
$D^0_{\m_{[p+1]}}$.

As $a_k$ has ghost number zero, the antighost number of $a_k$  should match its pureghost number. Consequently,  as the ghosts $A^{(p-q,q)}_{\m_{[q]}}$ and $D^0_{\m_{[p+1]}}$ have $ puregh=p$ and $q$ respectively, the equation $k=np+mq$ should be satisfied for some positive integers $n$ and $m$.
If there is no couple of integers $n,m$ to match $k$, then no $a_k$ satisfying the  equations of the descent (\ref{descente}) can be constructed and $a_k$ thus vanishes.

In the sequel, we will consider the case where $n$ and $m$
satisfying $k=np+mq$ can be found and classify the different cases
according to the value of $n$ and $m$: (i) $n\geqslant 2$, (ii)
$n=1$, (iii)  $n=0$, $m>1$, and (iv) $n=0$, $m=1$. We will show
that the corresponding candidates $a_k$  are either obstructed in
the lift to $a_0$ or that they are trivial, except in the case (iv).
In this case, $a_k$ can be lifted but $a_0$ depends explicitly
on $x$ and contains more than two derivatives.

\paragraph{(i) Candidates with $n\geqslant 2$ :}

The constraints $k\leqslant p+1$ and $k=np+mq$ have no solutions\footnote{There is
a solution in the case previously considered in  \cite{Boulanger:2000rq}, where $p=q=1$, $n=2$.
As shown in \cite{Boulanger:2000rq}, this solution gives rise to Einstein's theory of gravity.}.

\paragraph{(ii) Candidates with ${n=1}$ :}

The conditions $k=mq+p\leqslant p+1$ are only satisfied for $q=1=m$. As shown in \cite{Bekaert:2002uh}, the lift of these candidates is obstructed after one step without any additionnal assumption.

\paragraph{(iii) Candidates with $n=0$, ${m>1}$ :}

For a non-trivial candidate to exist at $k=mq$, Theorem \ref{thm8.1} tells us that $p$ and $q$ should
satisfy the relation $p+1=mq+s(q+1)$  for  some positive or null
integer $s$. The candidate then  has the form \bqn a^{D}_{mq} =
 C_{mq \, \n_{[q]}}^{*\,D-p-1+mq}\,
\o^{\n_{[q]}}_{(s,m)}(K,D)\,, \nonumber \eqn where what is meant
by a polynomial $\o_{(s,m)}$ is explained in Section
\ref{descentgammamodd}.
 \vspace*{.2cm}

We will  show that these candidates are either trivial or that there is an obstruction to lift them up to $a_0^D$ after $q$ steps.
\vspace*{.2cm}

It is straightforward to check that, for $1\leqslant  j \leqslant
q$, the terms
$$a^{D}_{mq-j} = C_{mq-j}^{*\, D-p-1+mq-j} \o^{s(q+1)+j,\,mq-j}$$
 satisfy the descent equations, 
since, as $m>1$, all antifields $C_{mq-j}^{*\, D-p-1+mq-j}$ are
invariant. The set of summed indices $\n_{[q]}$ is implicit as
well as the homogeneity degree of the generating polynomials
$\o_{(s,m)}$. We can thus lift $a^{D}_{mq}$ up to $a^D_{(m-1)q}$.
As $m>1$, this is not yet $a_0\,$.

There is however no $a^{D}_{(m-1)q-1}$ such that\bqn\g
(a^{D}_{(m-1)q-1}) +\d a^{D}_{(m-1)q} +d
\b^{D-1}_{(m-1)q-1}=0\,.\label{lookfor}\eqn Indeed, we have 
\bqn \d a^{D}_{(m-1)q}&=&-\g(C_{(m-1)q-1}^{*\,D-(s+1)(q+1)}\,
\o^{(s+1)(q+1),\,(m-1)q-1})
\nonumber \\
&&+(-)^{D-mq}\,C_{(m-1)q-1}^{*\,D-(s+1)(q+1)}
K^{q+1}\Big[\frac{\partial^L \o}{\partial D}\Big]^{s(q+1),(m-1)q}
\,. \nonumber \eqn  Without loss of generality, we can suppose
that
$$a^D_{(m-1)q-1} = C^{*\,D-(s+1)(q+1)}_{(m-1)q-1} \,\bar{a}^{(s+1)(q+1)}_0
+ \bar{a}^D_{(m-1)q-1}\,, $$ where there is an implicit summation
over all possible coefficients $\bar{a}^{(s+1)(q+1)}_0$, and most
importantly the two $\bar{a}$'s \textit{ do not}\footnote{This is
not true in the case
--- excluded in this paper --- where $p=q=1$ and $m=2$\,: since
$C^{*}_{(m-1)q-1}\equiv C^*_0 $ has antighost number zero, the antighost number counting  does not forbid that the $\bar{a}$'s
  depend on $C^*_0 $. Candidates arising in this way are treated in
\cite{Boulanger:2000ni} and give rise to a consistent deformation
of Fierz-Pauli's theory in $D=3$.} depend on $C^{*}_{(m-1)q-1} $.
Taking the Euler-Lagrange derivative of (\ref{lookfor}) with
respect to $C^{*}_{(m-1)q-1} $ yields \bqn \g
(\tilde{a}^{(s+1)(q+1)}_{0 }-\o^{(s+1)(q+1),\,(m-1)q-1}) \propto K^{q+1}\Big[\frac{\partial^L \o}{\partial
D}\Big]^{s(q+1),(m-1)q} \,. \nonumber \eqn    The product of
non-trivial elements of $H(\g)$ in the r.h.s. is not $\g$-exact
and constitutes an obstruction to the lift of the candidate,
unless it vanishes. 
The latter happens only when the polynomial
$\o_{(s,m)}$ can be expressed as $$\o^{\n_{[q]}}_{(s,m)}(K,D)=K^{q+1\,\m_{[p+1]}}\frac{\partial^L \tilde{\o}^{\n_{[q]}}_{(s-1,m+1)}(K,D)}{\partial D^{ \m_{[p+1]}}}\,,$$ for some
polynomial $\tilde{\o}^{\n_{[q]}}_{(s-1,m+1)}(K,D)$ of order $s-1$ in $K^{q+1}$ and $m+1$ in $D$.
 However, in this case,  $a^D_{mq}$ can be
removed by the trivial redefinition  $$a^D\rightarrow a^D+ s (\tilde{H}_{\n_{[q]}}\tilde{\o}^{\n_{[q]}}_{(s-1,m+1)} \vert ^D)\,.$$

This completes the proof that these candidates are either trivial
or that their lift is obstructed. As a consequence, they do not
lead to  consistent interactions and can be  rejected. Let us
stress that no extra assumption are needed to get this result. In the particular case $q=1$, this had already been  guessed  but not been proved in \cite{Bekaert:2002uh}.

\paragraph{(iv) Candidates with $n=0\,$, $m=1$ :}
These candidates exist only when the condition $p+2=(s+1)(q+1)$ is
satisfied, for some strictly positive integer $s\,$. It is
useful for the analysis to write the indices explicitly:
 \bqn
a^D_{q}&=&g^{\n_{[q]}\parallel\,\m^1_{[p+1]}\vert\, \ldots \vert\,
\m^{s+1}_{[p+1]}} \,C_{q\,\n_{[q]}}^{*\,D-p-1+q}  \left(
\prod_{i=1}^sK^{q+1}_{\m^i_{[p+1]}}\right)
D^{0}_{\m^{s+1}_{[p+1]}} \,, \nonumber \eqn where $g$ is a
constant tensor.

We can split the analysis into two cases: (i)  $g \rightarrow
(-)^q g$  under the exchange $\m^s_{[p+1]} \leftrightarrow
\m^{s+1}_{[p+1]}$, and (ii)  $g \rightarrow (-)^{q+1} g$ under the
same transformation.

In the case (i), $a_q^D$ can be removed by adding the trivial term
$s\, m^D$ where $m^D=\sum_{j=q}^{2q}m^D_j$ and $$
m^D_j=(-)^{D-q}\frac{1}{2} \,
g^{\n_{[q]}\parallel\,\m^1_{[p+1]}\vert\, \ldots \vert\,
\m^{s+1}_{[p+1]}} \,\, C^{*\,D-p-1+j}_{j\;\n{[q]}}\left(
\prod_{i=1}^{s-1}K^{q+1}_{\m^i_{[p+1]}}\right)  \Big[D_{\m_{[p+1]}^s}
D_{\m_{[p+1]}^{s+1}}\Big]^{2q+1-j}\,. $$ This construction
does not work in the case (ii) where the symmetry of $g$  makes
$m^D$ vanish.

In the case (ii), the candidate $a_q^D$ can be lifted up to $a_0^D$:
\bqn a^D_0\propto \,
f^{\s_{[p+1]}\parallel\,\m^1_{[p+1]}\vert\, \ldots \vert\,
\m^{s+1}_{[p+1]}} _{\t_{[D-p-q-1]}} x^{\t_1} dx^{\t_2} \ldots
dx^{\t_{D-p-q-1}}\,K^{q+1}_{\s_{[p+1]}}\,\left(
\prod_{i=1}^sK^{q+1}_{\m^i_{[p+1]}}\right) D^q_{\m^{s+1}_{[p+1]}}
\,,\nonumber \eqn where the constant tensor $f$ is defined by
$$f_{\t_{[D-p-q-1]}}^{\s_{[p+1]}\parallel\,\m^1_{[p+1]}\vert\, \ldots
\vert\, \m^{s+1}_{[p+1]}}\equiv
g^{\n_{[q]}\parallel\,\m^1_{[p+1]}\vert\, \ldots \vert\,
\m^{s+1}_{[p+1]}}\,\,\epsilon^{\s_{[p+1]}}_{\quad\,\,\,\,\,\,\n_{[q]}\t_{[D-p-q-1]}}\,.$$Let
us first note that this deformation does not affect the gauge
algebra, since it is linear in the ghosts.

The Lagrangian deformation $a_0^D$ depends explicitly on $x$,
which is not a contradiction with translation-invariance of the
physical theory if the $x$-dependance of the Lagrangian can be
removed by adding a total derivative and/or a $\d$-exact term. If
it were the case, $a_0^D$ would have the form $a_0^D=xG (\ldots) +
x^{\a}d (\ldots)_{\a}$. We have no complete proof that $a_0^D$
does not have this form, but it is not obvious and we think it
very unlikely. In any case, this deformation  is ruled out by the
requirement that the deformation of the Lagrangian contains at
most two derivatives. \vspace*{.2cm}

To summarize the results obtained in this section, we have proved
that, under the hypothesis of translation-invariance of the
first-order vertex $a^D_0$, all $a_k^D$ ($k>1$) can be removed by
trivial redefinitions of $a$, except when $p+2=(s+1)(q+1)$ for
some positive integer $s$. In that case, the supplementary
assumption that the deformed Lagrangian contains no more than two
derivatives is needed to reach the same conclusion, and the only possible deformation (without the latter
assumption) does not modify the gauge algebra.

\subsection{Computation of $a_1$}

The term $a_1$ vanishes without any further assumption
 when $q>1\,$. Indeed, when $q>1\,$,  the vanishing of the
cohomology of $\g$ in {\it puregh $1$} implies that there is
no non-trivial $a_1\,$.

This is not true when $q=1$, as there are some non-trivial
cocycles with pureghost number equal to one. However, it can
be shown \cite{Bekaert:2002uh} that any non-trivial $a_1^D$ leads
to a deformation of the Lagrangian with at least four derivatives.

\subsection{Computation of $a_0$}

This leaves us with the problem of solving the equation
$\g a_0^D+ d\, b_0^{D-1} = 0$ for $a_0^D\,$.
Such solutions correspond to
deformations of the Lagrangian that are invariant up to a total
derivative. Proceeding as in \cite{Boulanger:2004rx} and asking
for Lorentz invariance and that $a_0^D$ should not contain more than
two derivatives leaves only\footnote{ When $p=q$, there exists also a cosmological-like
term \cite{Boulanger:2004rx}:
$a_0=\Lambda \h_{\m_1 \n_1} \ldots \h_{\m_p \n_p}\phi^{\m_1 \ldots \m_p \vert\, \n_1 \ldots \n_p}\,.$} the Lagrangian itself. This deformation is of course trivial.

\section{Conclusions}
\label{sec:conclusions}

Assembling the results of the present paper ($p\neq q$) with those
previously obtained in \cite{Boulanger:2004rx } ($p=q\neq 1$), we
can state general conclusions for $[p,q]$-tensor gauge fields
where $p$ and $q$ are now arbitrary but both different from one.
Under the hypothesis of locality and translation invariance, there
is no smooth deformation of the free theory that modifies the
gauge algebra, which remains Abelian. This result strengthens the
conclusions of \cite{Bekaert:2002uh}, as no condition on the
number of derivative is needed any longer. Furthermore, for
$q>1\,$, when there is no positive integer $s$ such that
$p+2=(s+1)(q+1)\,$, there exists also no smooth deformation that
alters the gauge transformations. Finally, if one excludes
deformations that involve more than two derivatives in the
Lagrangian and are not Lorentz-invariant, then the only smooth
deformation of the free theory is a cosmological-like term for
$p=q$\cite{Boulanger:2004rx}.

These no-go results complete the search for self-interactions of
$[p,q]$-tensor gauge fields. It is still an open question whether
interactions are possible between $N$ different $[p,q]$-type fields (where ``different'' means $[p_1,q_1] \neq [p_2,q_2]$ for $N=2$),
or
with other types of fields.

As a conclusion, one can reformulate the results in more
physical terms by saying that no analogue of Yang-Mills nor
Einstein theories seems to exist for more exotic fields (at least
not in the range of local perturbative theories).


\section*{Acknowledgements}

We are grateful to M. Henneaux for proposing the project and for
numerous discussions. G. Barnich is also acknowledged for his
advices.

The work of X.B. is supported by the European Commission RTN
program HPRN-CT-00131, the one of N.B. is supported by a
Wiener-Anspach fellowship (Belgium), while the work of S.C. is
supported in part by the ``Actions de Recherche Concert{\'e}es''
of the ``Direction de la Recherche Scientifique - Communaut{\'e}
Fran{\c c}aise de Belgique'', by a ``P\^ole d'Attraction
Interuniversitaire'' (Belgium), by IISN-Belgium (convention
4.4505.86) and by the European Commission RTN program
HPRN-CT-00131, in which she is associated to K. U. Leuven.


\section*{Appendices}
\appendix

\section{Going to the Light-cone}
\label{BargmannWigner}

\begin{theorem}
\label{BWD} \noindent Let $K$ be a tensor in the irreducible
representation $[p+1,q+1]$ of $O(D-1,1)$. The space of such
{\textsl{harmonic multiforms}} $K$, \ie solutions of
$$
\left. \begin{array}{ll}\quad
    \pa_{[\m_0}K_{\m_1\ldots\m_{p+1}]\vert \n_1\ldots\n_{q+1}}=0=
    K_{\m_1\ldots\m_{p+1}\vert [\n_1\ldots\n_{q+1},\n_0]}
& ~~~{(\mbox{closed})} \\
            \quad
 \pa^{\m_1}K_{\m_1\ldots\m_{p+1}\vert \n_1\ldots\n_{q+1}}=0=
    \pa^{\n_1}K_{\m_1\ldots\m_{p+1}\vert \n_1\ldots\n_{q+1}}
& ~~~{(\mbox{coclosed})}
        \end{array}
\right\} \Longrightarrow  ~~ \Box K = 0\,
$$
is a unitary irreducible module of $O(D-2)$ associated to the
Young diagram $[p,q]$.
\end{theorem}

\noindent {\bf Proof : }\hspace*{.5pt}
Since $\Box K(x)=0$ then, after Fourier transform, $K(p)\neq
0$ iff $p^2=0\,$. In the light-cone frame, the light-like momentum
$p^{\m}$ decomposes into
$$
p^{\m}=(p^+,p^-,\;\underbrace{0,\ldots,0}_{D-2}\;)\,,~~ p^-=0\,.
$$
\begin{itemize}
\item[(i)] {
The condition that $K$ is closed implies
\bqn
\left\{ \begin{array}{ll}
p_{\m}\ve^{\m\n_1\ldots\n_{D-p-2}\m_1\ldots\m_{p+1}}
K_{\m_1\ldots\m_{p+1}\vert\a_1\ldots\a_{q+1}}=0\,, &\\
p_{\m}\ve^{\m\n_1\ldots\n_{D-q-2}\m_1\ldots\m_{q+1}}
K_{\a_1\ldots\a_{p+1}\vert \m_1\ldots\m_{q+1}}=0\,, &
\end{array}\right.
\nonumber
\eqn
\ie
\bqn
\left\{ \begin{array}{ll}
\ve^{-\n_1\ldots\n_{D-p-2}\m_1\ldots\m_{p+1}}
K_{\m_1\ldots\m_{p+1}\vert\a_1\ldots\a_{q+1}}=0\,, &\\
\ve^{-\n_1\ldots\n_{D-q-2}\m_1\ldots\m_{q+1}}
K_{\a_1\ldots\a_{p+1}\vert \m_1\ldots\m_{q+1}}=0\,. &
\end{array}\right.
\nonumber \eqn The latin indices will run over the $D-2$
transverse values. Assigning $\n_1=+\,$, $\n_2=j_2\,$, $\cdots\;$,
$\n_{D-\ell-2}=j_{D-\ell-2}\,$ (where $\ell=p$ or $q$
respectively), one finds
$$
K_{i_1\ldots i_{p+1}\vert\a_1\ldots\a_{q+1}}=0=
K_{\a_1\ldots\a_{p+1}\vert i_1\ldots i_{q+1}}\,.
$$
In other words, $K$ vanishes whenever one of its columns contains only
transverse indices.
}
\item[(ii)] {
The fact that $K$ is coclosed on-shell implies 
$$
p^+ K_{+\m_2\ldots\m_{p+1}\vert\a_1\ldots\a_{q+1}}=0=
p^+ K_{\a_1\ldots\a_{p+1}\vert +\m_2\ldots\m_{q+1}} \,,
$$
\ie
$$
K_{+\m_2\ldots\m_{p+1}\vert\a_1\ldots\a_{q+1}}=0=
K_{\a_1\ldots\a_{p+1}\vert +\m_2\ldots\m_{q+1}}\,.
$$
In other words, $K$ vanishes whenever one of its columns contains a
``$+$'' index.
}
\end{itemize}
Once it has been observed that each column of $K$ must contain at
least one ``$-$'' index and no ``+'' index, one finds that the
tensor
$$
\phi_{i_1\ldots i_{p}\vert j_1\ldots j_{q}}
\equiv
\frac{(p+1)(q+1)}{p_{-}^2} K_{-i_1\ldots i_{p}\vert j_1\ldots j_{q}-}
$$
obeys\bqn
0&=&\frac{p+2}{p_{-}^2}K_{[-i_1\ldots i_{p}\vert j_1]\ldots j_q -}
= \phi_{[i_1\ldots i_{p}\vert j_1]\ldots j_q}\,,
\nonumber \\
0&=&\h^{\m_1\n_1}
K_{\m_1\m_2\ldots\m_{p+1}\vert\n_1\ldots\n_{q+1}}
\nonumber \\
&\Rightarrow&0=\frac{(p+1)(q+1)}{p_{-}^2}\,\d^{i_1j_1}
K_{-i_1i_2\ldots i_{p}\vert j_1\ldots j_{q}-}=\d^{i_1j_1}
\phi_{i_1i_2\cdots i_{p}\vert j_1\cdots j_{q}}\,. \nonumber \eqn
\qedsymbol

\section{Proof of Theorem \ref{BK}}
\label{ProofoftheoremBK}

In this appendix, we give the proof of Theorem \ref{BK}:
\vspace{.2cm}

{\it
Let $r$ be a strictly positive integer. A complete set of
representatives of $H_k^D(\d \vert\, d)$ ($k>1$ and $k=D-r(D-p-1)\geqslant q $ ) is
given by the terms of form-degree $D$ in all homogeneous
polynomials $P^{(r)}(\tilde{H})$ of degree $r$ in $\tilde{H}$ (or equivalently $P(\tilde{\cal H})$ of degree $r$ in $\tilde{\cal H}$).
}\vspace{.4cm}

It is obvious from the definition of $\tilde{H}$ and from equation
(\ref{htilde}) that the term of form-degree $D$ in $P^{(r)}(\tilde{H})$
has the right antighost number and is a cocycle of $H_k^D(\d
\vert\, d)$. 
Furthermore, as $\tilde{\cal H}=\tilde{H}+ d( \ldots)$, $P^{(r)}(\tilde{\cal H})$ belongs to the same cohomology class as $P^{(r)}(\tilde{H})$ and can as well be chosen as a representative of this class.
To prove the theorem, it is then enough, by Theorem
\ref{BKzero}, to prove that the cocycle $P^{(r)}(\tilde{H})\vert \,_k^D$ is non-trivial. The proof
is by induction:  we know the theorem to be true for $r=1$ by
Theorem \ref{pplusun}, supposing that the theorem is true for $r-1$,
(\ie  $[P^{(r-1)}(\tilde{H})]^D_{k+D-p-1}$ is not trivial in
$H^D_{k+D-p-1}(\d\vert d)$) we prove that $[P^{(r)}(\tilde{H})]^D_k$
is not trivial either.
\vspace{.2cm}

Let us assume that
$[P^{(r)}(\tilde{H})]^D_k$ is trivial: $[P^{(r)}(\tilde{H})]^D_k= \d (u_{k+1} d^Dx) + d v_k^{D-1}$.
We  take the Euler-Lagrange derivative of this equation with respect to $C^{*}_{k,\m_{[q]}\vert \n_{[p+1-k]}}$.
For $k> q$, it reads:
\bqn
\a_{\m_{[q]} \vert\, \nu_{[p+1-k]}} =(-)^k \d(Z_{1\; \m_{[q]} \vert\, \nu_{[p+1-k]}})-Z_{0 \;\m_{[q]} \vert\, [\nu_{[p-k]},\n_{p+1-k}]} \,, \label{eul}
\eqn
where
\bqn
\a_{\m_{[q]} \vert\, \nu_{[p+1-k]}} d^D x &\equiv &\frac{\d^L [P^{(r)}(\tilde{H})]^D_k}{\d C_k^{*\; \m_{[q]} \vert\,\nu_{[p+1-k]}}}\,, \nonumber \\
Z_{k+1-j \; \m_{[q]} \vert\, \nu_{[p+1-j]}} &\equiv& \frac{\d^L u_{k+1}}{\d C_j^{*\; \m_{[q]}
 \vert\, \nu_{[p+1-j]}}} \,, \, \;{\rm for} \;  j =k, k+1 \,.\nonumber
\eqn
For $k=q$, there is an additional term:
\bqn
\a_{\m_{[q]} \vert\, \nu_{[p+1-q]}} =(-)^q \d(Z_{1\; \m_{[q]} \vert\, \nu_{[p+1-q]}})-(Z_{0 \;\m_{[q]} \vert\, [\nu_{[p-q]},\n_{p+1-q}]} - Z_{0 \;[\m_{[q]} \vert\, \nu_{[p-q]},\n_{p+1-q}]} )\,.\label{eul2}
\eqn
The origin of the additional term lies in the fact that
$C_q^{*\; \m_{[q]} \vert\,  \nu_{[p+1-q]}}$
does not possess all the irreducible components of $[q] \otimes [p+1-q]\,$: the completely antisymmetric component $[p+1]$ is missing. Taking the  Euler-Lagrange derivative with respect to this field thus involves projecting out this component.

We will first solve the equation (\ref{eul}) for $k>q$, then come back to  (\ref{eul2}) for $k=q$.
\vspace{.2cm}

Explicit computation of  $\a_{\m_{[q]} \vert\, \nu_{[p+1-k]}} $ for $k>q$ yields:
\bqn
\a_{\m_{[q]} \vert\, \nu_{[p+1-k]}}
=
[\tilde{H}^{\r^1_{[q]}}]_{0,\,\s^1_{[D-p-1]}}
\ldots [\tilde{H}^{\r^{r-1}_{[q]}}]_{0,\,\s^{r-1}_{[D-p-1]}}
a_{\m_{[q]} \vert\r^1_{[q]} \vert \ldots \vert\r^{r-1}_{[q]}}\d^{[\s^1_{[D-p-1]}\ldots \s^{r-1}_{[D-p-1]}]}_{\nu_{[p+1-k]}} 
\nonumber  \,,
\eqn
where $a$ is a constant tensor and the notation  $[A]_{k,\,\n_{[p]}}$ means the coefficient $A_{k,\,\n_{[p]}}$, with antighost number $k$, of the $p$-form component of $A=\sum_{k,l} A_{k,\,\n_{[l]}}dx^{\n_1} \ldots dx^{\n_{l}}$.
Considering the indices $\nu_{[p+1-k]}$ as form indices, (\ref{eul}) reads:
\bqn
\a^{p+1-k}_{\m_{[q]}}&=& [\tilde{H}^{\r^1_{[q]}}]_0^{D-p-1} \ldots [\tilde{H}^{\r^{r-1}_{[q]}}]_0^{D-p-1}a_{\m_{[q]} \vert \r^1_{[q]} \vert \ldots \vert\r^{r-1}_{[q]}}=\Big[\prod_{i=1}^{(r-1)}\tilde{H}^{\r^{i}_{[q]}}\Big]^{p+1-k}_{0}
a_{\m_{[q]} \vert\r^1_{[q]} \vert \ldots \vert\r^{r-1}_{[q]}}
\nonumber \\
&=&
(-)^k \d(Z^{p+1-k}_{1\; \m_{[q]} })+(-)^{p-k+1} d\,Z^{p-k}_{0 \;\m_{[q]} }\,.\nonumber
\eqn
The latter equation is equivalent to $$ \Big[\prod_{i=1}^{(r-1)}\tilde{H}^{\r^{i}_{[q]}}\Big]^{D}_{D-p-1+k} a_{\m_{[q]} \vert \r^1_{[q]} \vert \ldots \vert\r^{r-1}_{[q]}}=\d (\ldots) + d (\ldots)\,,$$
which contradicts the induction hypothesis. The assumption that
$[P^{(r)}(\tilde{H})]^D_k$ is trivial is thus wrong, which proves the theorem for $k>q$.
\vspace{.2cm}

The  philosophy of the
 resolution of (\ref{eul2}) for $k=q$ is inspired by the proof of Theorem 3.3 in \cite{Boulanger:2004rx} and goes as follows: first, one has to constrain the last term of (\ref{eul2}) in order to get an equation similar to the equation
(\ref{eul}) treated previously,
then one solves this equation in the same way as for $k>q$.

Let us constrain the last term of (\ref{eul2}). Equation (\ref{eul2}) and explicit computation of $\a_{\m_{[q]} \vert\, \nu_{[p+1-k]}}$ imply
\bqn
\pa_{[\n_{p+1-q}}\a_{\m_{[q]} \vert\,
\n_{[p-q]}] \l}&=&(-)^q \d(\pa_{[\n_{p+1-q}}Z_{1 \; \m_{[q]} \vert\,
\n_{[p-q]}] \l} ) - b \,\pa_{[\n_{p+1-q}}Z_{0 \; \m_{[q]}
\vert\, \nu_{[p-q]}],\l}\hspace{2.5cm}
\nonumber \\
&\approx &\!\!b \,\pa_{\l} (
[\tilde{H}^{\r^1_{[q]}}]_{0,\,\s^1_{[D-p-1]}} \ldots [\tilde{H}^{\r^{r-1}_{[q]} }]_{0 ,\,\s^{r-1}_{[D-p-1]}} \d^{[\s^1_{[D-p-1]}\ldots \s^{r-1}_{[D-p-1]}]}_{[\nu_{[p+1-k]}}a_{\m_{[q]} ]\vert\r^1_{[q]} \vert \ldots \vert\r^{r-1}_{[q]}})
\nonumber
\eqn where
$b=\frac{q}{(p+1)(p+1-q)} $.
By the isomorphism $H_0^0(d\vert \d)/\mathbb{R}\cong H^D_{D}(\d\vert d)\cong 0\,$, the latter equation implies
\bqn
Z_{0 \; [\m_{[q]}
\vert\, \nu_{[p-q]},\n_{p+1-q}]}\approx -[\tilde{H}^{\r^1_{[q]}}]_{0,\,\s^1_{[D-p-1]}} \ldots [\tilde{H}^{\r^{r-1}_{[q]} }]_{0,\,\s^{r-1}_{[D-p-1]}} a_{\m_{[q]} \vert\r^1_{[q]} \vert \ldots \vert\r^{r-1}_{[q]}}\d^{[\s^1_{[D-p-1]}\ldots \s^{r-1}_{[D-p-1]}]}_{\nu_{[p+1-k]}}\nonumber
\eqn
(the constant solutions are removed by considering the equation in polynomial degree $r-1$ in the fields and antifields.).
Inserting this expression for $Z_{0 \; [\m_{[q]}
\vert\, \nu_{[p-q]},\n_{p+1-q}]}$ into (\ref{eul2}) and redefining $Z_1$ in a suitable way yields (\ref{eul}) for $k=q$. The remaining of the proof is then the same as for $k>q$. \hspace{.5cm}\qedsymbol

\section{Proof of Theorem \ref{cohoinv}}
\label{Proofoftheoremcohoinv}

In this appendix, we give the complete (and lengthy) proof of
Theorem \ref{cohoinv}: \vspace{.2cm}

{\it Let $a^D_k$ be an invariant polynomial. If $a_k^D = \d
b_{k+1}^D + d c_k^{D-1} $, then
$$a_k^D = P_{(s,r)}(K^{q+1},\tilde{\cal H})\vert \,_k^D+\d \m_{k+1}^D + d \n_k^{D-1}\,,$$
where  $P_{(s,r)}(K^{q+1},\tilde{\cal H})$ is a polynomial of degree $s$ in $K^{q+1}$ and $r$ in $\tilde{\cal H}$, such that the integers $s,r\geqslant 1$ satisfy
 $D=r(D-p-1)+k+s(q+1)$ and $\m_{k+1}^D$ and $\n_k^{D-1}$ are invariant polynomials.
}

The proof is by induction and follows closely the steps of the
proof of similar theorems in the case of $1$-forms
\cite{Barnich:db,Barnich:mt},
$p$-forms\cite{Henneaux:1996ws}, gravity
\cite{Boulanger:2000rq} or $[p,p]$-fields \cite{Boulanger:2004rx}.

There is a general procedure to prove that the theorem \ref{cohoinv} holds for $k>D$,
that can be found e.g. in \cite{Boulanger:2000rq} and will not be repeated here.
We assume that the theorem  has been proved for any $k^{'}>k$, and show that it is
still valid for $k\,$.
\vspace*{.2cm}

The proof of the induction step is rather lengthy and is
decomposed into several steps:
\begin{itemize}
\item
the Euler-Lagrange derivatives of $a_k$ with respect to the fields
$\phi$ and $C^*_j$ ($1\leqslant j \leqslant p+1$) are computed in terms of
the Euler-Lagrange derivatives of $b_{k+1}$ (section
\ref{sec7.1});
\item
it is shown that the Euler-Lagrange derivatives of $b_{k+1}$ can
be replaced by invariant quantities in the expression for the
Euler-Lagrange derivative of $a_k$ with the lowest antighost
number, up to some additionnal terms (section \ref{sec7.2});
\item
 the previous step is extended to all the Euler-Lagrange derivatives of $a_k$
(section \ref{sec7.3});
\item
the Euler-Lagrange derivative of $a_k$ with respect to the field
$\phi$ is reexpressed in terms of invariant quantities (section
\ref{sec7.4});
\item
an homotopy formula is used to reconstruct $a_k$ from its
Euler-Lagrange derivatives (section \ref{sec7.5}).
\end{itemize}

\subsection{Euler-Lagrange derivatives of $a_k$}
\label{sec7.1}

We define
\bqn
Z_{k+1-j\; \m_{[q]} \vert\, \nu_{[p+1-j]}} &=& \frac{\d^L b_{k+1}}{\d C_j^{* \; \m_{[q]}
 \vert\, \nu_{[p+1-j]}}} \,, \quad 1 \leqslant j \leqslant p+1 \,,\nonumber \\
Y_{k+1}^{\m_{[p]} \vert\,  \nu_{[q]}} &= &\frac{\d^L b_{k+1}}{\d
\phi_{\m_{[p]} \vert\,  \nu_{[q]}}} \,.\nonumber
\eqn
Then, the Euler-Lagrange derivatives of $a_k$ are given by
\bqn \frac{\d^L a_k}{\d C_{p+1}^{*\; \m_{[q]} }}
 &=&(-)^{p+1} \d Z_{k-p \; \m_{[q]}} \,,
 \label{vroum} \\
\frac{\d^L a_k}{\d C_j^{*\; \m_{[q]} \vert\, \nu_{[p+1-j]}}}
&=&(-)^j \d Z_{k+1-j \; \m_{[q]} \vert\, \nu_{[p+1-j]}} -Z_{k-j
\; \m_{[q]}\vert\, [\nu_{[p-j]},\n_{p+1-j}]} \;,\,q<j \leqslant p\,,
\nonumber\\
\frac{\d^L a_k}{\d C_j^{*\; \m_{[q]} \vert\, \nu_{[p+1-j]}}}
&=&(-)^j \d Z_{k+1-j \; \m_{[q]} \vert\, \nu_{[p+1-j]}} -Z_{k-j
\; \m_{[q]}\vert\, [\nu_{[p-j]},\n_{p+1-j}]} \vert\,_{sym\, of\,
C^{*}_j}\;,1\leqslant j \leqslant q \,,\hspace{1cm}
\nonumber\\
\frac{\d^L a_k}{\d \phi^{\m_{[p]} \vert\,  \nu_{[q]}}} &=&\d
Y_{k+1 \; \m_{[p]} \vert\,  \nu_{[q]}} + \b D_{\m_{[p]} \vert\,
\nu_{[q]} \vert\, \r_{[p]} \vert\, \s_{[q]} }Z_k^{\s_{[q]}
\vert\,\r_{[p]} }\,, \label{EL}
\eqn
where $\b\equiv(-)^{(q+1)(p+\frac{q}{2})} \frac{(p+1)!}{q!(p-q+1)!}\,$, and
$D^{\m_{[p]}\vert \hspace{1.2cm}\s_{[q]} }_{\hspace{.6cm}\nu_{[q]} \vert\, \r_{[p]} \vert}\equiv \frac{1}{(p+1)!q!}\;\delta^{[ \s_{[q]}   \a \m_{[p]}
]}_{[\n_{[q]}\b  \r_{[p]}]} \pa_{\a}\pa^{ \b}$ is the
second-order self-adjoint differential operator defined by
$G_{\m_{[p]} \vert\, \nu_{[q]}}\equiv$
$D_{\m_{[p]} \vert\, \nu_{[q]} \vert\, \r_{[p]} \vert\, \s_{[q]}}$
$C^{\r_{[p]} \vert\, \s_{[q]} }\,$.

As in Appendix \ref{ProofoftheoremBK}, the projection on the symmetry of the indices of $C^*_j$ is needed when $j\leqslant q$, since in that case the variables
$C^*_j$ do not possess all the irreducible components of $[q] \otimes [p+1-j]\,$,
but only those where the length of the first column is smaller or equal to $p\,$. When $j>q$, the projection is trivial.

\subsection{Replacing $Z$ by an invariant in the Euler-Lagrange derivative of $a_k$ with
the lowest antighost number}
\label{sec7.2}

We should first note that, when $k<p+1\,$, some of the
Euler-Lagrange derivatives of $a_k$ vanish identically: indeed, as
there is no negative antighost-number field, $a_k$ cannot depend
on $C^{*}_j$ if $j>k$. Some terms on the r.h.s. of
(\ref{vroum})-(\ref{EL}) also vanish:  $Z_{k+1-j}$ vanishes when
$j>k+1\,$. This implies that the $p+1-k$ top  equations of
(\ref{vroum})-(\ref{EL}) are trivially satisfied: the $p-k$ first
equations involve only vanishing terms, and the $(p-k+1)$th
involves in addition the $\d$ of an antighost-zero term, which
also vanishes trivially.
The first non-trivial equation is then \bqn \frac{\d^L a_k}{\d
C^{*}_{k \; \m_{[q]} \vert\, \nu_{[p+1-k]}}} &=&(-)^k \d(Z_{1\;
\m_{[q]} \vert\, \nu_{[p+1-k]}} )-Z_{0 \; \m_{[q]} \vert\,
[\nu_{[p-k]},\n_{p+1-k}]} \vert\,_{sym\, of\, C^{*}_k} \,.
\label{ping}\eqn

Let us now define $[T^q_{\r_{[p+1]}}]_{\n_{[q]}}\equiv (-)^{q} \pa_{[\r_1}\phi_{\r_2 \ldots \r_{p+1}] \vert \n_{[q]}}$. We will prove the following lemma for $k\geqslant q\,$:
\begin{lemma} \label{nontriv}
In the first non-trivial equation of the system
(\ref{vroum})-(\ref{EL}) (\ie (\ref{vroum}) when
$k\geqslant p+1$ and (\ref{ping}) when $p+1 > k \geqslant q$), 
 respectively $Z_{k-p}$ or $Z_1$  
satisfies
\bqn 
Z_{l\;\m_{[q]} \vert\, \nu_{[p+l-k]}}
&=&Z^{\prime}_{l\; \m_{[q]} \vert\, \nu_{[p+l-k]}}+(-)^{k-l}\d \b_{l+1\; \m_{[q]} \vert\, \nu_{[p+l-k]}} + \b_{l\; \m_{[q]}  \vert\,
[ \nu_{[p+l-k-1]},\nu_{p+l-k} ]}\vert\,_{sym\, of\, C^{*}_{k-l+1}} \nonumber \\
&+&A_l \Big[P^{(n)}_{\m_{[q]}}(\tilde{\cal H})+\frac{1}{s}T^q_{\r_{[p+1]}} \frac{\pa^L R^{(s,r)}_{\m_{[q]}}(K^{q+1},\tilde{\cal H}) }{\pa K^{q+1}_{\r_{[p+1]}}}\Big]_{l,\,\nu_{[p+l-k]}}\vert\,_{sym\, of\, C^{*}_{k-l+1}}  
,
\hspace{.8cm}\label{basis}\eqn 
where $Z^{\prime}_l$ is invariant, the
$\b_l$'s are at least linear in ${\cal N}$ and possess the same
symmetry of indices as $Z_{l-1}\,$, $A_l\equiv(-)^{lp+p+1+\frac{l(l+1)}{2}}\,$,
$P^{(n)}$ is a polynomial of degree $n$ in $\tilde{\cal H}$ and $R^{(s,r)}$ is a polynomial of degree $s$ in $K^{q+1}$ and $r$ in $\tilde{\cal H}$. The polynomials are present only when $p-k=n(D-p-1)$ or $p+1-k=s(q+1)+r(D-p-1)$ respectively. 
\vspace{.1cm}

Moreover, when $p+1 > k \geqslant q$,  the first non-trivial equation can be written
\bqn \frac{\d^L a_k}{\d
C^{*}_{k \; \m_{[q]} \vert\, \nu_{[p+1-k]}}} &=&(-)^{k} \d
Z_{1\,\m_{[q]}\vert\, \nu_{[p+1-k]}}^{\prime}-Z_{0\,\m_{[q]}\vert\, [\nu_{[p-k]},\n_{p+1-k}]}^{\prime}\vert\,_{sym\, of\, C^{*}_{k}}  \nonumber \\
&&+\Big([Q^{(m)}_{\m_{[q]}}(K^{q+1}) ]_{\nu_{[p+1-k]}}
+(-)^k [R^{(s,r)}_{\m_{[q]}}(K^{q+1},\tilde{\cal H}) ]_{0,\,\nu_{[p+1-k]}}\Big)\vert\,_{sym\, of\, C^{*}_{k}} \,, 
\nonumber \eqn
where $Z_{0}^{\prime}$ is an invariant  and 
$Q^{(m)}_{\m_{[q]}}(K^{q+1}) $ is a polynomial of degree $m$ in $K^{q+1}$, present only when $p+1-k=m(q+1)$.
\end {lemma}

The lemma will be proved in the sections
\ref{inductbasis1}--\ref{inductbasis3} respectively for the cases
$k \geqslant p+1\,$, $ q < k <p+1$ and $k=q\,$.

\subsubsection{Proof of Lemma \ref{nontriv} for $k \geqslant p+1$}
\label{inductbasis1}

As $k-p>0\,$, there is  no trivially satisfied equation  and we
start with the top equation of (\ref{vroum})--(\ref{EL}).

The  lemma \ref{nontriv} is a direct consequence of the well-known  Lemma \ref{preliminary}
(see e.g. \cite{Boulanger:2000rq} ):
\begin{lemma}\label{preliminary}Let $\a$ be an invariant local form that is
$\d$-exact, \ie $\a=\d\b\,$. Then $\b=\b^\prime+\d\s\,$, where
$\b^\prime$ is invariant and we can assume without loss of
generality that $\s$ is at least linear in the variables of $\cal N\,$.
\end{lemma}

\subsubsection{Proof of Lemma \ref{nontriv} for $q < k <  p+1$}

The first non-trivial equation is (as $k>q\,$):
\bqn
\frac{\d^L a_k}{\d C^{*}_{k \; \m_{[q]} \vert\,
\nu_{[p+1-k]}}} &=&(-)^k \d(Z_{1\; \m_{[q]} \vert\, \nu_{[p+1-k]}}
)-Z_{0 \;\m_{[q]} \vert\, [\nu_{[p-k]},\n_{p+1-k}]} \,.
\label{pingg}
\eqn
We will first prove that $Z_1$ has the required
form, then we will prove the the first non-trivial equation can indeed be reexpressed as stated in Lemma \ref{nontriv}.

\paragraph{First part:}
Defining
$\a_{0\;\m_{[q]}\vert \nu_{[p+1-k]}}\equiv\frac{\d^L a_q}{\d
C^{*}_{q \; \m_{[q]} \vert\, \nu_{[p+1-q]}}}$, the above equation can be
written  as \be \a_0^{p+1-k}=(-)^k \d (Z_1^{p+1-k})+ (-)^{p+1-k}
dZ_0^{p-k} \,,\label{alpha} \ee where we consider the indices $\nu_{[p+1-k]}$ as form-indices and omit to write the indices $\m_{[q]} $. Acting with $d$ on this equation
yields $d\a_0^{p+1-k}=(-)^{k+1} \d (dZ_1^{p+1-k})$. Due to Lemma
\ref{preliminary}, this implies that \be \a_1^{p+2-k} =
dZ_1^{p+1-k}+ \d Z_2^{p+2-k}\,,\label{beta}\ee for some invariant
$\a_1^{p+2-k}$ and some $Z_2^{p+2-k}$. These steps can be
reproduced to build a descent of equations ending with
\bqn\a_{D-p-1+k}^{D} = dZ_{D-p-1+k}^{D-1}+ \d
Z_{D-p+k}^{D}\,,\nonumber \eqn
where $\a_{D-p-1+k}^{D} $ is invariant.
As  
$D-p-1+k>k$, the induction hypothesis can be used and 
implies
$$\a_{D-p-1+k}^{D} =
dZ_{D-p-1+k}^{\prime\,\,D-1}+ \d Z_{D-p+k}^{\prime\,\,D}+[R(K^{q+1},\tilde{\cal H})]^D_{D-p-1+k}\,,$$
where $Z_{D-p+k}^{\prime\,\,D}$ and $Z_{D-p-1+k}^{\prime\,\,D-1}$ are invariant, and
$R(K^{q+1},\tilde{\cal H})$ is a polynomial of order $s$ in $K^{q+1}$ and $r$ in $\tilde{\cal H}$ (with $r,s>0$), present when $p+1-k=s(q+1)+r(D-p-1)$.
This equation can be lifted and implies that
$$\a_1^{p+2-k} = dZ_1^{\prime\, p+1-k}+ \d Z_2^{\prime \, p+2-k}+[R(K^{q+1},\tilde{\cal H})]^{p+2-k}_{1}\,,$$
for some invariant quantities $Z_1^{\prime\, p+1-k}$ and $Z_2^{\prime \,
p+2-k}$. 
Substracting the last equation from (\ref{beta}) yields
 $$
d\Big(Z_1^{p+1-k}-Z_1^{\prime\, p+1-k} -\frac{1}{s}T^q \Big[\frac{\pa^L R(K^{q+1},\tilde{\cal H})}{\pa K^{q+1}}\Big]_1^{p+1-k-q}\Big)+\d(\ldots)= 0\,.$$
As $H_1^{p+1-k}(d \vert\, \d)\cong H_{D-(p-k)}^{D} (\d \vert\,d)$, by Theorem \ref{BK} the solution of this equation is
\bqn Z_1^{p+1-k}=Z_1^{\prime\, p+1-k}
+\frac{1}{s}T^q \Big[\frac{\pa^L R(K^{q+1},\tilde{\cal H})}{\pa K^{q+1}}\Big]_1^{p+1-k-q}+d \b_1^{p-k} + \d \b_2^{ p+1-k} +[P^{(n)}(\tilde{\cal H})]_1^{p+1-k}  \,,\nonumber\eqn
where the last term is present only when $p-k=n(D-p-1)$. 

This proves the first part of the induction basis, regarding $Z_1$.

\paragraph{Second part:}
We insert
the above result for $Z_1$ into (\ref{alpha}). Knowing that $\d(
[P(\tilde{\cal H})]_1^{p+1-k} )+ d (
[P(\tilde{\cal H})]_0^{p-k} )=0\,$
and defining $$W_0^{p-k}=(-)^{k+1} \Big((-)^{p} Z_0^{p-k}+ \d
\b_1^{p-k} + [P^{(n)}(\tilde{\cal H})]_0^{p-k} 
+\frac{1}{s}T^q \Big[\frac{\pa^L R(K^{q+1},\tilde{\cal H})}{\pa K^{q+1}}\Big]_0^{p-k-q} 
\Big)\,,$$  we get
\bqn
\a_0^{p+1-k}= (-)^{k} \d(Z_1^{\prime\,p+1-k})+d(W_0^{p-k})
+(-)^{k}[R(K^{q+1},\tilde{\cal H})]_0^{p-k}
\,. \nonumber
\eqn
Thus $d(W_0^{p-k})$ is an invariant and the invariant Poincar\'e
Lemma \ref{invPoinclemma} then states that
$$d(W_0^{p-k})=d(Z_0^{\prime\;p-k})+Q(K^{q+1}) $$
for some invariant $Z_0^{\prime\;p-k}$ and some polynomial in $K^{q+1}$, $Q(K^{q+1})$.
This straightforwardly implies 
\bqn \a_0^{p+1-k}=(-)^{k} \d
(Z_1^{\prime\,p+1-k})+d(Z_0^{\prime\;p-k})+Q(K^{q+1}) +(-)^{k}[R(K^{q+1},\tilde{\cal H})]_0^{p-k}\,, \nonumber \eqn
which completes the proof of Lemma \ref{nontriv} for $q < k <  p+1$.
 $\qedsymbol $

\subsubsection{Proof of Lemma \ref{nontriv} for $k=q$}
\label{inductbasis3}

The first non-trivial equation is \bqn \frac{\d^L a_q}{\d C^{*}_{q
\; \m_{[q]} \vert\, \nu_{[p+1-q]}}} =(-)^q \d(Z_{1 \;  \m_{[q]}
\vert\, \nu_{[p+1-q]}} ) -(Z_{0 \; \m_{[q]} \vert\,
[\nu_{[p-q]},\n_{p+1-q}]}  - Z_{0 \; [\m_{[q]} \vert\,
\nu_{[p-q]},\n_{p+1-q}]} )\,.\label{indbasq} \eqn
 This equation is different from the equations treated in the previous cases because the operator acting on $Z_0$ cannot
be seen as a total derivative, since it involves the projection on a specific Young diagram.
The latter problem was already faced in the $[p,p]$-case and the philosophy of the
 resolution goes as follows \cite{Boulanger:2004rx}:
\begin{itemize}
\item[(1)] one first constrains the last term of (\ref{indbasq}) to get an equation similar to Equation
(\ref{ping}) treated previously,
\item[(2)]  one solves it in the same way as for $q<k<p+1\,$.
\end{itemize}

We need the useful lemma \ref{lem1}, proved in  \cite{Boulanger:2004rx}.
\begin{lemma}\label{lem1}
If $\a_0^1$ is an invariant polynomial of antighost number 0 and
form degree 1 that satisfies $\alpha_{0 }^1 = \delta Z_{1}^1 +
d W_{0}^0 \,, $ then, for some invariant
polynomials ${Z'}_{1}^1$ and ${W'}_{0}^0\,$, $ Z_1^1 ={Z'}_1^1+\d \phi_2^1 + d \chi_{1}^0$ and $
W_{0}^0= {W'}_{0}^0 + \d  \chi_1^0\,. $
\end{lemma}

As explained above, we now constrain the last term of
(\ref{indbasq}). Equation
  (\ref{indbasq}) implies
\bqn \pa_{[\r}\a_{0\;\m_{[q]} \vert\,
\n_{[p-q]}]\nu_{p+1-q}}=(-)^q \d(\pa_{[\r}Z_{1 \; \m_{[q]} \vert\,
\n_{[p-q]}] \nu_{p+1-q}} ) - b \,\pa_{[\r}Z_{0 \; \m_{[q]}
\vert\, \nu_{[p-q]}],\n_{p+1-q}}\,, \nonumber \eqn where
$b\equiv\frac{q}{(p+1)(p+1-q)} $. Defining
\bqn
\tilde{\a}^1_{0\,[\r\m_{[q ]} \n_{[p-q]}]}&=&\pa_{[\r}\a_{0\;\m_{[q ]} \vert\, \n_{[p-q]}]
\nu_{p+1-q}} dx^{\n_{p+1-q}}\,, \nonumber \\
\tilde{Z}^1_{1\, [\r\m_{[q ]} \n_{[p-q]}]}&=&(-)^q \pa_{[\r}Z_{1 \; \m_{[q ]} \vert\,
\n_{[p-q]}] \nu_{p+1-q}} dx^{\n_{p+1-q}}\,, \nonumber \\
\tilde{W}^0_{0\,[\r\m_{[q ]} \n_{[p-q]}]}&=&- a\,\pa_{[\r}Z_{0 \;
\m_{[q ]} \vert\,  \nu_{[p-q]}]}\,, \nonumber
\eqn
and omitting to write the indices $[\r\m_{[q]}\n_{[p-q]}]$, the above equation reads $
\tilde{\a}^1_0=\d\tilde{Z}^1_1+ d\tilde{W}^0_0 \,$. Lemma
\ref{lem1} then implies that $\tilde{W}^0_0=I^{\prime\,0}_{0} + \d
m^0_1$ for some invariant $I^{\prime\,0}_{0}$. By the definition of
$\tilde{W}^0_0$, this statement is equivalent to \bqn
\pa_{[\r}Z_{0 \; \m_{[q]} \vert\,
\nu_{[p-q]}]}=I^{\prime}_{0\,[\m_{[q]} \nu_{[p-q]}\r]}+ \d
m_{1\,[\m_{[q]} \n_{[p-q]}\r]} \,.\nonumber \eqn Inserting this result into
(\ref{indbasq}) yields \bqn \a_{0\;\m_{[q]}\vert\,
\nu_{[p+1-q]}}-I^{\prime}_{0\,[\m_{[q]}  \nu_{[p+1-q]}]}=\d( (-)^q Z_{1
\; \m_{[q]} \vert\, \nu_{[p+1-q]}} +m_{1\,[\m_{[q]}
\nu_{[p+1-q]}]}) -Z_{0 \; \m_{[q]}  \vert\,
[\nu_{[p-q]},\n_{p+1-q}]}\,.\nonumber \eqn This equation has the same
form as (\ref{pingg}) and can be solved in the same way to get the
following result: \bqn Z_{1 \; \m_{[q]}\vert\,
\n_{[p+1-q]}}&=&(-)^{q+1} m_{1 \; [\m_{[q]}\n_{[p+1-q]}]}+
Z^{\prime}_{1 \; \m_{[q]}\vert\, \n_{[p+1-q]}}
+\b_{1\;\m_{[q]}\vert\, [ \nu_{[p-q]},\n_{p+1-q}]}+\d \b_{2\;\m_{[q]}\vert\, \n_{[p+1-q]}}
\nonumber\\
&&+\frac{1}{s}\Big[T^q_{\r_{[p+1]} }\frac{\pa^L R_{\m_{[q]}}(K^{q+1},\tilde{\cal H})}{\pa K^{q+1}_{\r_{[p+1]} }}\Big]_{1,\, \n_{[p+1-q]}}+[P(\tilde{\cal H})]_{1,\,\n_{[p+1-k]}}\,,
\nonumber\\
\a_{0\;\m_{[q]}\vert\, \n_{[p+1-q]}}&=&I^{\prime}_{0\,[\m_{[q]}\vert\,
\n_{[p+1-q]}]}+(-)^q \d (Z^{\prime}_{1 \; \m_{[q]}\vert\,
\n_{[p+1-q]}})+Z_{0 \; \m_{[q]}\vert\,
[\nu_{[p-q]},\n_{p+1-q}]}^{\prime}
\nonumber \\
&&+ [Q_{\m_{[q]}}(K^{q+1})]_{\n_{[p+1-q]}}+(-)^{k}[R(K^{q+1},\tilde{\cal H})]_{0,\,\n_{[p+1-q]}}\,.\nonumber \eqn
Removing the completely antisymmetric parts of these equations
yields the desired result. \qedsymbol \vspace*{.2cm}

This ends the proof of Lemma \ref{nontriv} for $k\geqslant q\,$.

\subsection{Replacing all $ Z$ and $Y$ by invariants}
\label{sec7.3}

We will now prove the following lemma:
\begin{lemma} \label{yzinv}
The Euler-Lagrange derivatives of $a_k$ can be written
\bqn
\frac{\d^L a_k}{\d C_{p+1}^{* \;  \m_{[q]} }} &=&(-)^{p+1} \d(Z^{\prime}_{k-p \; \m_{[q]}  } )\,,
\nonumber\\
\frac{\d^L a_k}{\d C_j^{*\; \m_{[q]}  \vert\, \nu_{[p+1-j]}}} &=&(-)^j \d(Z^{\prime}_{k+1-j \;
\m_{[q]}  \vert\, \nu_{[p+1-j]}} )-Z^{\prime}_{k-j \; \m_{[q]}  \vert\, [ \nu_{[p-j]},\n_{p+1-j}]}
\;\;, q<j \leqslant p \,,\nonumber\\
\frac{\d^L a_k}{\d C_j^{* \; \m_{[q]}  \vert\,  \nu_{[p+1-j]}}} &=&(-)^j \d(Z^{\prime}_{k+1-j \;
\m_{[q]} \vert\,  \nu_{[p+1-j]}} )-Z^{\prime}_{k-j \; \m_{[q]}  \vert\, [ \nu_{[p-j]},\n_{p+1-j}]}
\vert\,_{sym\, of\, C^{*}_j}\;,1\leqslant j \leqslant q \,,\nonumber\\
\frac{\d^L a_k}{\d \phi^{\m_{[q]}\vert\, \nu_{[q]}}} &=&\d
(Y^{\prime}_{k+1 \; \m_{[q]}\vert\, \nu_{[q]}} )+ \b D_{\m_{[q]}\vert\,
\nu_{[q]} \vert\,
 \r_{[p]} \vert\, \s_{[q]} }{Z'}_k^{\s_{[q]}
\vert\,\r_{[p]}  }\,, \nonumber\eqn where $Z_{l}^{\prime}$
($k-p\leqslant l \leqslant k$)  and $Y_{k+1}^{\prime}$ are invariant polynomials,
except in the following cases. When $k=p+1-m(q+1)$ for some strictly positive integer $m\,$, there is an additionnal term in the first non-trivial
equation:
\bqn
\frac{\d^L a_{k}}{\d C_k^{*\; \m_{[q]}\vert\,  \nu_{[p+1-k]}}}=(-)^k \d Z^{\prime}_{1\;
\m_{[q]}\vert\,  \nu_{[p+1-k]}} -Z^{\prime}_{0 \; \m_{[q]}\vert\, [\nu_{[p-k]},\n_{p+1-k}]}
+[Q_{\m_{[q]}}(K^{q+1})]_{\nu_{[p+1-k]}}\vert\,_{sym\, of\, C^{*}_k}
\,,\nonumber \eqn
where $Q$ is a polynomial of degree $m$ in $K^{q+1}$.
Furthermore, when $k=p+1-r(D-p-1)-s(q+1)$ for a couple of integer $r,s>0$, then there is an additional term in each Euler-Lagrange derivative:
\bqn
\frac{\d^L a_k}{\d C_j^{* \; \m_{[q]}  \vert\,  \nu_{[p+1-j]}}} &=&(-)^j \d(Z^{\prime}_{k+1-j \;
\m_{[q]} \vert\,  \nu_{[p+1-j]}} )-Z^{\prime}_{k-j \; \m_{[q]}  \vert\, [ \nu_{[p-j]},\n_{p+1-j}]}
\vert\,_{sym\, of\, C^{*}_j}\nonumber\\
&&+ (-)^{k+p+1}A_{k-j}[R_{\m_{[q]}}(K^{q+1},\tilde{\cal H})]_{k-j\;\nu_{[p+1-j]}}\vert\,_{sym\, of\, C^{*}_j}
\nonumber\\
\frac{\d^L a_k}{\d \phi^{\m_{[q]}\vert\, \nu_{[q]}}} &=&\d
(Y^{\prime}_{k+1 \; \m_{[q]}\vert\, \nu_{[q]}} )+ \b D_{\m_{[q]}\vert\,
\nu_{[q]} \vert\,
 \r_{[p]} \vert\, \s_{[q]} }{Z'}_k^{\s_{[q]}
\vert\,\r_{[p]}  }\nonumber\\
&&+A\,\d^{[\s_{[q]} \a \m_{[p]}\xi]}_{[\n_{[q]}\b \r_{[p+1]}]}
\pa_\a \pa^\b (x_{\xi} \,[R_{\s_{[q]}}(K^{q+1},\tilde{\cal H})]_k^{\r_{[p+1]}})\,,
\nonumber
\eqn
where $A=\b \frac{p+q+2}{(D-p-q-1)(p+1)!q!}A_k(-)^{p+k+1}$.
\end{lemma}

\noindent {\bf{Proof:}}\hspace{.5cm}
By Lemma \ref{nontriv}, we know that  the $Z$'s
involved in the first non-trivial equation satisfy (\ref{basis})
and that this equation has the required form. We
will proceed by induction and prove that when $Z_{k-j}$  (where $k-j\geqslant 1$) satisfies
(\ref{basis}), then the equation for $\frac{\d^L a_k}{\d C^{*}_j}$ also has the desired form and $Z_{k-j+1}$ also satisfies (\ref{basis}). 

Let us assume that $Z_{k-j }$ satisfies  (\ref{basis}) and
consider the following equation:
\bqn
\frac{\d^L a_k}{\d C^{*}_{j\; \m_{[q]}\vert\, \n_{[p+1-j]} }}=(-)^j \d(Z_{k+1-j}^{
\m_{[q]}\vert\, \n_{[p+1-j]}} )-Z_{k-j }^{ \m_{[q]} \vert\,
[\nu_{[p-j]},\n_{p+1-j}]}\vert\,_{sym\, of\,
C^{*}_j}\,. \label{xunk}
\eqn
Inserting  (\ref{basis}) for $Z_{k-j}$ into this equation yields
\bqn
\frac{\d^L a_k}{\d C^{*}_{j\; \m_{[q]}\vert\, \n_{[p+1-j]}}}&=&(-)^j \d \Big( Z_{k+1-j }^{ \m_{[q]}\vert\, \n_{[p+1-j]} }-\b_{k-j+1}^{\m_{[q]}\vert\, [ \nu_{[p-j]}, \nu_{p-j+1}]}\vert\,_{sym\, of\,C^{*}_j} \Big)\label{eqtot}
\\ 
&+&(-)^{k+p}a_{k-j}\d \Big[P^{\m_{[q]}}(\tilde{\cal H})+\frac{1}{s}T^q_{\r_{[p+1]}}\frac{\pa^L R^{\m_{[q]}}(K^{q+1},\tilde{\cal H})}{\pa K^{q+1}_{\r_{[p+1]}}}\Big]_{k-j+1}^{\n_{[p+1-j]} }\vert\,_{sym\, of\, C^{*}_j}
\nonumber \\
&+& \Big(-Z_{k-j }^{\prime\; \m_{[q]}\vert\, [\nu_{[p-j]},\n_{p+1-j}]}+(-)^{p+k+1}A_{k-j}[R^{\m_{[q]}}(K^{q+1},\tilde{\cal H})]_{k-j}^{\n_{[p+1-j]}}\Big)\vert\,_{sym\, of\, C^{*}_j} 
\nonumber
\,.
\eqn
Note that one can omit to
project on the symmetries of $C^*_{j+1}$ when inserting
(\ref{basis}) into  (\ref{xunk}). Indeed the Young components that
are removed by this projection would be removed later anyway by
the projection on the symmetries of $C^*_j\,$.

Defining the invariant
\bqn Z_{k+1-j}^{\prime\; \m_{[q]}\vert\, \n_{[p+1-j]} }&\equiv&Z_{k+1-j }^{
\m_{[q]}\vert\, \n_{[p+1-j]} }\vert\,_{{\cal N}=0} \nonumber \\
&&+(-)^{k+p+j}A_{k-j} \Big[P^{\m_{[q]}}(\tilde{\cal H})+\frac{1}{s}T^q_{\r_{[p+1]}}\frac{\pa^L R^{\m_{[q]}}(K^{q+1},\tilde{\cal H})}{\pa K^{q+1}_{\r_{[p+1]}}}\Big]_{k-j+1}^{\n_{[p+1-j]} }\vert\,_{sym\, of\, C^{*}_j}\vert\,_{{\cal N}=0}
\nonumber\eqn
and setting ${\cal N}=0$ in the last equation  yields, as $\b_{k-j+1}$ is at least linear in ${\cal
N}$, \bqn \frac{\d^L a_k}{\d C^{*}_{j \; \m_{[q]}\vert\,
\n_{[p+1-j]}}} =(-)^j \d(Z_{k+1-j }^{\prime\; \m_{[q]}\vert\,
\n_{[p+1-j]}} )-Z_{k-j}^{\prime\; \m_{[q]}\vert\, [
\nu_{[p-j]},\n_{p+1-j}]} \vert\,_{sym\, of\, C^{*}_j}
\nonumber \\
+(-)^{p+k+1}A_{k-j}[R^{\m_{[q]}}(K^{q+1},\tilde{\cal H})]_{k-j}^{\n_{[p+1-j]}}\vert\,_{sym\, of\, C^{*}_j} 
\,.
\label{eqinv}
\eqn
This proves the part of the induction regarding the equations for the Euler-Lagrange derivatives. We now prove that $Z_{k-j+1}$ verifies (\ref{basis}).

Substracting (\ref{eqinv}) from
(\ref{eqtot}), we get
\bqn
0 \,&= &(-)^j \d  \Big(Z_{k+1-j }^{\m_{[q]}\vert\, \n_{[p+1-j]}}
-Z_{k+1-j }^{\prime\; \m_{[q]}\vert\, \n_{[p+1-j]}}-\b_{k+1-j}^{ \m_{[q]} \vert\, [ \nu_{[p-j]},
\nu_{p+1-j}]}\vert\,_{sym\, of\, C^{*}_j}\nonumber \\
& &\hspace{1cm}+(-)^{j+k+p}A_{k-j}\Big[P^{\m_{[q]}}(\tilde{\cal H})+\frac{1}{s}T^q_{\r_{[p+1]}}\frac{\pa^L R^{\m_{[q]}}(K^{q+1},\tilde{\cal H})}{\pa K^{q+1}_{\r_{[p+1]}}}\Big]_{k+1-j}^{\n_{[p+1-j]} }\vert\,_{sym\, of\, C^{*}_j}\Big)\,. \nonumber
\eqn

As $k+1-j>0$, this implies \bqn Z_{k+1-j }^{ \m_{[q]}\vert\,
\n_{[p+1-j]}}&=&Z_{k+1-j }^{\prime\; \m_{[q]}\vert\, \n_{[p+1-j]}}+
(-)^{j-1} \d \b_{k-j}^{ \m_{[q]}\vert\, \n_{[p+1-j]}}+\b_{k-j+1}^{
\m_{[q]}\vert\, [\nu_{[p-j]}, \nu_{p+1-j}]}\vert\,_{sym\, of\,
C^{*}_j}
\nonumber \\
&& +A_{k+1-j}\Big[P^{\m_{[q]}}(\tilde{\cal H})+\frac{1}{s}T^q_{\r_{[p+1]}}\frac{\pa^L R^{\m_{[q]}}(K^{q+1},\tilde{\cal H})}{\pa K^{q+1}_{\r_{[p+1]}}}\Big]_{k+1-j}^{\n_{[p+1-j]} }\vert\,_{sym\, of\, C^{*}_j} \,,\nonumber
\eqn
which is the expression (\ref{basis}) for $Z_{k+1-j }$.

Assuming that $Z_{k-j}$ satisfies  (\ref{basis}) , we have thus proved
that the equation for $\frac{\d^L a_k}{\d C_j^{*}} $ has the
desired form and that $Z_{k+1-j}$ also satisfies (\ref{basis}).
Iterating this step, one shows that all $Z$'s satisfy
(\ref{basis}) and that the equations involving only $Z$'s have the
desired form.

It remains to be proved that the Euler-Lagrange derivative with respect to the field takes the right form. Inserting the expression (\ref{basis}) for $Z_k$ into (\ref{EL}) and some algebra yield
\bqn
\frac{\d^L a_k}{\d \phi^{\m_{[q]}\vert\, \nu_{[q]}}} &=&\d
(\tilde{Y}_{k+1 \; \m_{[q]}\vert\, \nu_{[q]}} \vert\,_{sym\, of\,
\phi})+ \b D_{\m_{[q]}\vert\,
\nu_{[q]} \vert\,
 \r_{[p]} \vert\, \s_{[q]} }{Z'}_k^{\s_{[q]}
\vert\,\r_{[p]}  }\nonumber\\
&&+A\,\d^{[\s_{[q]} \a \m_{[p]}\xi]}_{[\n_{[q]}\b \r_{[p+1]}]}
\pa_\a \pa^\b (x_{\xi} \,[R_{\s_{[q]}}(K^{q+1},\tilde{\cal H})]_k^{\r_{[p+1]}})\vert\,_{sym\, of\,
\phi}\,,
\nonumber
\eqn
where \bqn\tilde{Y}_{k+1 \; \m_{[q]}\vert\, \nu_{[q]}} &\equiv & Y_{k+1 \; \m_{[q]}\vert\, \nu_{[q]}}
+ \b D_{\m_{[q]}\vert\,\nu_{[q]} \vert\,\r_{[p]} \vert\, \s_{[q]} }\b_{k+1}^{\s_{[q]}\vert\, \r_{[p]}}
\nonumber \\
&&+\,c \,\d^{[\s_{[q]}\a \m_{[p]} ]}_{[\n_{[q]}\b \r_{[p]}]} \pa_{\a
} \Big[P_{\s_{[q]}}(\tilde{\cal H})+ \frac{1}{s} T^q_{\l_{[p+1]}}\frac{\pa^L R^{\s_{[q]}}(K^{q+1},\tilde{\cal H})}{\pa K^{q+1}_{\l_{[p+1]}}}\Big]_{k+1}^{[\r_{[p]}\b]}\nonumber \\
&&+(-)^{k+q+1}A\,\d^{[\s_{[q]} \a \m_{[p]}\xi]}_{[\n_{[q]}\b \r_{[p+1]}]}
\pa_\a (x_{\xi} \, [R_{\s_{[q]}}(K^{q+1},\tilde{\cal H})]_{k+1}^{[\r_{[p+1]} \b]})
\nonumber \eqn
and  $c\equiv\b \frac{1}{(p+1)!q!}A_k (-)^{p+k+1}$. Defining $Y_{k+1\; \m_{[p]} \vert\, \n_{[q]}}^{\prime}\equiv \tilde{Y}_{k+1 \; \m_{[q]}\vert\, \nu_{[q]}} \vert\,_{sym\, of\,
\phi}\vert_{{\cal N}=0}$ and setting ${\cal N}=0$ in the above equation completes the proof of Lemma \ref{yzinv}. \qedsymbol

\subsection{Euler-Lagrange derivative with respect to the field}
\label{rewriting}
\label{sec7.4}

In this section, we manipulate the Euler-Lagrange derivative of
$a_k$ with respect to the field $\phi\,$.

We have proved in the previous section that it can be written in
the form \bqn \frac{\d^L a_k}{\d \phi^{\m_{[p]} \vert\,
\n_{[q]}}} &=&\d (Y^{'}_{k+1 \; \m_{[p]} \vert\, \n_{[q]}} )+ \b
D_{\m_{[p]} \vert\, \n_{[q]} \vert\,  \r_{[p]} \vert\,  \s_{[q]}
}Z_k^{'\,\s_{[q]}\vert\,\r_{[p]} }\nonumber\\
&&+A\,\d^{[\s_{[q]} \a \m_{[p]}\xi]}_{[\n_{[q]}\b \r_{[p+1]}]}
\pa_\a \pa^\b (x_{\xi} \,[R_{\s_{[q]}}(K^{q+1},\tilde{\cal H})]_k^{\r_{[p+1]}})\vert\,_{sym\, of\,
\phi}\,.\nonumber \eqn 
As $a_k$
is invariant, it can depend on $\phi_{\m_{[p]} \vert\, \n_{[q]}}$
only through $K_{\m_{[p]} \a\vert\, \n_{[q]} \b}$, which implies
that $ \frac{\d^L a_k}{\d \phi^{\m_{[p]} \vert\, \n_{[q]}}}
=\pa^{\a \b}X_{[\m_{[p]} \a]\vert\, [\n_{[q]} \b]}\,,$ where $X$ has the symmetry of the curvature. This
in turn implies that $\d(Y^{'}_{k+1 \; \m_{[p]} \vert\, \n_{[q]}}
)=\pa^{\a \b}W_{\m_{[p]}  \a\vert\, \n_{[q]} \b}$ for some $W$
with the Young symmetry $  [p+1,q+1]\,$.
Let us consider the indices $ \m_{[p]} $ as form indices. As
$H_{k+1}^{D-p} (\d \vert\, d) \cong H_{p+1+k}^{D}(\d \vert\,
d)\cong 0 $ for $k>0$, the last equation implies \bqn Y^{'}_{k+1
\; \m_{[p]} \vert\, \n_{[q]}} = \d A_{k+2 \; \m_{[p]} \vert\,
\n_{[q]}}+ \pa^{\l}T_{k+1 \; [\l \m_{[p]} ] \vert\, \n_{[q]}}\,.
\label{yprime} \eqn By the induction hypothesis  for $p+1+k$ , we
can take $A_{k+2}$ and $T_{k+1}$ invariant. Antisymmetrizing
(\ref{yprime}) over the indices $\m_q \ldots \m_p \nu_1 \ldots
\nu_{q}$ yields \bqn 0=\d A_{k+2 \; \m_1 \ldots \m_{q-1}[\m_q
\ldots \m_p \vert\, \nu_1 \ldots \nu_{q}]}+ \pa^{\l}T_{k+1 \; \l
\m_1 \ldots \m_{q-1}[\m_q \ldots \m_p \vert\, \nu_1 \ldots
\nu_{q}]}\,. \nonumber \eqn The solution of this equation for
$T_{k+1}$ 
is 
\bqn T_{k+1\, \m_0 \ldots
\m_{q-1}[\m_q \ldots \m_p \vert\, \nu_1 \ldots \nu_{q}]}= \d
Q_{k+2 \, \m_0 \ldots \m_{q-1}\vert\, [\m_q \ldots \m_p \nu_1
\ldots \nu_{q}]}+ \pa^{\a}S_{k+1 \, \a\m_0 \ldots
\m_{q-1}\vert\, [\m_q \ldots \m_p \nu_1 \ldots
\nu_{q}]}
\nonumber \\
+\Big[U_{[\m_q \ldots \m_p \nu_1 \ldots \nu_{q}]}^{(u)}(\tilde{\cal H})\Big]_{k+1}^{\r_{[D-q]}} \epsilon_{\m_0 \ldots \m_{q-1}\r_{[D-q]}}
\,,\nonumber\eqn
where $U^{(u)}$ is a polynomial of degree $u$ in $\tilde{\cal H}$, present when $k+q+1=D-u(D-p-1)$ for some strictly positive integer $u$.
As $T$ and $U^{(u)}(\tilde{\cal H})$ are invariant, we can use the induction hypothesis for $k^{\prime}=k+1+q$. This implies
\bqn
T_{k+1\, \m_0 \ldots
\m_{q-1}[\m_q \ldots \m_p \vert\, \nu_1 \ldots \nu_{q}]}&=& \d
Q^{\prime}_{k+2 \,  \m_0 \ldots \m_{q-1}\vert\, [\m_q \ldots \m_p \nu_1
\ldots \nu_{q}]}+ \pa^{\a}S^{\prime}_{k+1 \, \a\m_0 \ldots
\m_{q-1}\vert\, [\m_q \ldots \m_p \nu_1 \ldots
\nu_{q}]}\label{S}
\\
&+&\Big[U_{[\m_q \ldots \m_p \nu_1 \ldots \nu_{q}]}^{(u)}(\tilde{\cal H})+V_{[\m_q \ldots \m_p \nu_1 \ldots \nu_{q}]}^{(v,w)}(K^{q+1},\tilde{\cal H})\Big]_{k+1}^{\r_{[D-q]}} \epsilon_{\m_0 \ldots \m_{q-1}\r_{[D-q]}}\,,
\nonumber 
\eqn
where $Q_{k+2 }^{\prime}$ and $S_{k+1 }^{ \prime}$ are invariants and 
$V^{(v,w)}$ is a polynomial of order $v$ and $w$ in $K^{q+1}$ and $\tilde{\cal H}$ respectively, present when $D-q=v(q+1)+w(D-p-1)+k+1$ for some strictly positive integers $v,w$. 
\vspace*{.2cm}

We define the invariant tensor $E_{\a \m_{[p]} \vert\, \b  \nu_{[q]} }$
with Young symmetry $  [p+1,q+1]$ by \bqn E_{\a \m_{[p]} \vert\,
\b \nu_{[q]} }= \sum_{i=0}^{q+1} \a_i S^{\prime}_{k+1\;\r_0 \ldots \r_{i-1}
[ \n_i \ldots \n_q \vert\, \b \n_1 \ldots \n_{i-1}] \r_i \ldots
\r_p} \d^{[\r_0 \ldots \r_p]}_{[\a \m_{[p]}] } \nonumber\eqn
where $\a_i =  \a_0 \frac{(q+1)!}{(q+1-i)! \, i! }$ and $\a_0 = (-)^{pq}
\frac{((p+1)!)^2}{ (p-q)!\, (q!)^2 \, (p-q+1) \,(p+2)\, \sum_{j=0}^{q}\frac{(p-j)!}{(q-j)!}}\,$.
\vspace*{.2cm}

Writing $\pa^{\a \b}E_{k+1\; \a \m_{[p]} \vert\, \b \nu_{[q]} }$
in terms of $S^{ \prime}_{k+1}$ and using (\ref{S}) and (\ref{yprime})
yields \bqn Y^{'}_{k+1 \;  \m_{[p]} \vert\, \nu_{[q]}}&=&\pa^{\a
\b}E_{k+1\; \a \m_{[p]} \vert\, \b \nu_{[q]} }+\d F_{k+2 \;
\m_{[p]} \vert\, \nu_{[q]}} \nonumber \\
&&+\pa^{\a} \sum_{i=0}^{q} \b_i \Big[V_{[\a \n_{[i]}\m_{i+1} \ldots \m_p]}^{(v,w)}(K^{q+1},\tilde{\cal H})\Big]_{k+1}^{\r_{[D-q]}} \epsilon_{\m_{[i]}\n_{i+1} \ldots \n_q\r_{[D-q]}}\,,\label{yprime2}
\eqn where $F_{k+2}$ is invariant, $\b_i \equiv \a_0 \frac{(p+2) q!}{(p+1)\, i! \,
(q-i)!}$ and $v$ is allowed to take the value $v=0$ to cover also the case of the polynomial $U^{(w)}(\tilde{\cal H})$.

\subsection{Homotopy formula}
\label{sec7.5}

We will now use the homotopy formula to reconstruct $a_k$ from its
Euler-Lagrange derivatives: \bqn a^D_k
&=&\int_0^1dt \Big[ \phi_{
\m_{[p]} \vert\,  \nu_{[q]}}\frac{\d^L a_k}{\d \phi_{\m_{[p]}
\vert\, \nu_{[q]}}} + \sum_{j=1}^{p+1} C^{*}_{j \;  \m_{[q]}
\vert\, \nu_{[p+1-j]}}\frac{\d^L a_k}{\d C^{*}_{j \; \m_{[q]}
\vert\, \nu_{[p+1-j]}}} \Big]\, d^Dx\,.
\nonumber \eqn
Inserting the expressions for the Euler-Lagrange derivatives given by Lemma \ref{yzinv} yields
\bqn
a^D_k &=&\int_0^1dt \Big[\d(\phi_{\m_{[p]}
\vert\,  \nu_{[q]}}\,Y_{k+1} ^{\prime\; \m_{[p]} \vert\,  \nu_{[q]}})
+\sum_{j=1}^{p+1} \d (C^{*}_{j \;  \m_{[q]} \vert\, \nu_{[p+1-j]}}Z_{k+1-j}^{\prime\;\m_{[q]}
\vert\,  \nu_{[p+1-j]}}) 
\nonumber \\
&&\hspace{1cm}+\sum_{j=1}^{k} C^{*}_{j \;  \m_{[q]} \vert\, \nu_{[p+1-j]}}(-)^{k+p+1}A_{k-j}[R^{\m_{[q]} }(K^{q+1},\tilde{\cal H})]_{k-j}^{\nu_{[p+1-j]}}
\nonumber \\
&&\hspace{1cm}+\phi_{\m_{[p]}
\vert\,  \nu_{[q]}}A\,\d^{[\s_{[q]} \a \m_{[p]}\xi]}_{[\n_{[q]}\b \r_{[p+1]}]}
\pa_\a \pa^\b (x_{\xi} \,[R_{\s_{[q]}}(K^{q+1},\tilde{\cal H})]_k^{\r_{[p+1]}})
\nonumber \\
&&\hspace{1cm}+\, C^{*}_{k\;  \m_{[q]} \vert\, \nu_{[p+1-k]}} [Q^{(m)\,\m_{[q]}}(K^{q+1})]^{\nu_{[p+1-k]}}\Big]  d^Dx 
+ d n^{D-1}_{k}\,.
\nonumber\eqn
Using the result  (\ref{yprime2}) for $Y_{k+1} ^{\prime}$ and some algebra, one finds
\bqn
a^D_k &=&\int_0^1dt \Big[\d (K_{\m_{[p+1]}\vert \n_{[q+1]}}E_{k+1}^{\m_{[p+1]}\vert \n_{[q+1]}}d^Dx)+ a_v \, K_{\m_{[p+1]}}^{q+1}[V^{(v,w)\,\m_{[p+1]}}(K^{q+1},\tilde{\cal H})]_k^{D-q-1}  
\nonumber \\
&&\hspace{1cm}+\sum_{j=1}^{p+1} \d (C^{*}_{j \;  \m_{[q]} \vert\, \nu_{[p+1-j]}}Z_{k+1-j}^{\prime\;\m_{[q]}
\vert\,  \nu_{[p+1-j]}}d^Dx) 
+a_r [\tilde{\cal H}^{\s_{[q]}}\,R_{\s_{[q]}}(K^{q+1},\tilde{\cal H})]_{k}^{D}
\nonumber \\
&&\hspace{1cm}+ a_q \, [\tilde{\cal H}^{\s_{[q]}}]_k^{D-m(q+1)} \, Q^{(m)}_{\s_{[q]}}(K^{q+1})\Big]+ d \bar{n}^{D-1}_{k}
\,,\nonumber
\eqn
where $a_v=(-)^{k(q+1)} \sum_{i=0}^{q} \b_i \frac{i!(p-i)!}{p!}\,$, 
$a_r= (-)^{D(p+k+1)+\frac{p(p+1)+k(k+1)}{2}}$
 and $a_q=(-)^k a_r\,$.
In short,
\bqn
a^D_k =\, [P(K^{q+1},\tilde{\cal H})]_k^D  +\d \m^D_{k+1} +d \bar{n}^{D-1}_{k}\nonumber
\eqn
for some invariant $\m^D_{k+1}$, and some polynomial $P$ of strictly positive order in $K^{q+1}$ and $\tilde{\cal H}$.
\vspace{.2cm}

We still have to prove that $\bar{n}^{D-1}_{k}$ can be taken
invariant. 

\noindent Acting with $\g$ on the last equation yields $d (\g
\bar{n}^{D-1}_{k}) =0$. By the Poincar\'e lemma, $\g
\bar{n}^{D-1}_{k}= d (r_k^{D-2})$. Furthermore,  a well-known
result on $H(\g\vert\, d)$ for positive antighost number $k$ (see e.g. Appendix A.1 of \cite{Boulanger:2000rq}) 
states that one can redefine $\bar{n}^{D-1}_{k}$ in such a way that $\g \bar{n}^{D-1}_{k}=0$. As the pureghost number of
$\bar{n}^{D-1}_{k}$ vanishes, the last equation implies that $\bar{n}^{D-1}_{k}$ is an
invariant polynomial.
\vspace{.2cm}

This completes the proof of Theorem \ref{cohoinv} for $k\geqslant q$.\hspace{.5cm}\qedsymbol

\end{document}